\documentstyle[12pt,psfig]{article}
\newcommand {\beq}{\begin{equation}}
\newcommand {\eeq}{\end{equation}}
\newcommand {\beqa}{\begin{eqnarray}}
\newcommand {\eeqa}{\end{eqnarray}}
\newcommand {\beqan}{\begin{eqnarray*}}
\newcommand {\eeqan}{\end{eqnarray*}}
\newcommand {\n}{\nonumber \\}

\newcommand {\Romannumeral}[1]{\uppercase\expandafter{\romannumeral#1}}

\newcommand {\tr} {\mbox{tr}}

\begin{document}
\setlength{\oddsidemargin}{0cm}
\setlength{\baselineskip}{7mm}  %7mm 

\begin{titlepage}
\renewcommand{\thefootnote}{\fnsymbol{footnote}}
  \begin{normalsize}
    \begin{flushright}
      DPNU-98-08\\
      UT-805\\
      hep-th/9802082
    \end{flushright}
  \end{normalsize}

  \begin{Large}
    \vspace{1.0cm}
    \begin{center}
      {\Large Numerical Study of the Double Scaling Limit \\
        in Two-Dimensional Large $N$ Reduced Model}\\
%SU($N$) reduced gauge theory} \\
    \end{center}
  \end{Large}

  \vspace{1cm}

\begin{center}
    Takayuki N{\sc akajima}$^{1)}$
\footnote{E-mail address : nakajima@siren.phys.s.u-tokyo.ac.jp}
%nakajima@theory.kek.jp}
{\sc and}
    Jun N{\sc ishimura}$^{2)}$
\footnote{E-mail address : nisimura@eken.phys.nagoya-u.ac.jp}\\
      \vspace{1cm}
$1)$       {\it Department of Physics, University of Tokyo,}\\
               {\it Bunkyo-ku, Tokyo 113, Japan}\\
%        $\ast$ {\it National Laboratory for High Energy Physics (KEK),}\\
%               {\it Tsukuba, Ibaraki 305, Japan}\\
$2)$       {\it Department of Physics, Nagoya University,} \\
              {\it Chikusa-ku, Nagoya 464-01, Japan}\\
%      \vspace{1cm}
\vspace{15mm}

\end{center}
\hspace{5cm}

  \begin{abstract}
\noindent
We study the two-dimensional Eguchi-Kawai model as a toy
model of the IIB matrix model, which has been recently proposed 
as a nonperturbative definition of the type IIB superstring theory.
While the planar limit of the model is known to reproduce the
two-dimensional Yang-Mills theory,
we find through Monte Carlo simulation that the model allows 
a different large $N$ limit, which
can be considered as the double scaling limit in matrix models.

\vspace{0.5cm}
{\it keywords:} string theory; large N gauge theory; 
matrix model; Monte Carlo simulation
\end{abstract}

\end{titlepage}

\vfil\eject

\setcounter{footnote}{0}

\section{Introduction}
\setcounter{equation}{0}

It is no doubt that
the nonperturbative study of string theories is one of the most exciting
topic in particle physics, since it might provide all the answers
to the fundamental questions concerning our world,
such as the space-time dimension, the gauge 
group, the three generations of the matter fields and so on.
Indeed the nonperturbative study of the bosonic string theory 
in less than one dimension was successfully done
through the double scaling limit of the matrix model \cite{DSL}
some time ago.
Much effort has been made towards the generalization of the approach
to bosonic strings in more than one dimensions 
\cite{morethan,ABC}
%,KawaiOkamoto} 
as well as to superstrings \cite{MikovicSiegel,MarinariParisi}.

Recently, a new type of matrix model has been proposed as a
nonperturbative definition of the superstring theory.
The model proposed by Banks, Fischler, Shenker and Susskind \cite{BFSS},
which is conjectured to provide a nonperturbative definition 
of the M-theory \cite{M-theory} in the infinite-momentum frame, is 
a matrix quantum mechanics, which can be obtained by dimensional reduction
of the 10D large $N$ super Yang-Mills theory to one dimension,
while the one proposed 
by Ishibashi, Kawai, Kitazawa and Tsuchiya \cite{IKKT},
which is conjectured to provide a nonperturbative definition 
of type IIB superstring theory,
is a matrix model, which can be obtained by dimensional reduction
of the 10D large $N$ super Yang-Mills theory to a point.
It is now conventional to call the former as M(-atrix) theory
and the latter as IIB matrix model.
Although these proposals are basically based on
the spirit to describe the strings in terms of matrices
of very large size, 
there are many novel features that didn't show up 
in the old-fashioned matrix models.
One of them is the appearance of the Yang-Mills theory.
This is related to the effective action \cite{Dbrane} of D-particles
and D-instantons, respectively for the two models.
It can also be viewed to be related to the supermembrane action 
and the Green-Schwartz action in the Schild gauge, respectively.
Another interesting feature of these models is that 
all the target-space coordinates come out as the eigenvalues of
the matrices.

For the IIB matrix model, 
the light-cone string field Hamiltonian of the type IIB superstring
theory has been reproduced \cite{FKKT}
by identifying the string field for fundamental strings 
with the Wilson loop and analyzing the Schwinger-Dyson equation.
As an important consequence of this study,
they have clarified how one should take the double scaling limit
in this model.
The IIB matrix model can be considered as an example of
large $N$ reduced models (See Ref. \cite{Das} for a review.), 
which have been studied so far
exclusively in the planar limit,
in which the models are equivalent to the field theory before being
reduced.
Whether a large $N$ reduced model allows 
any sensible double scaling limit is itself a nontrivial question.

In this paper, we study the two-dimensional Eguchi-Kawai model
\cite{EK}, as a toy model of the IIB matrix model.
The model is nothing but the SU($N$) lattice
gauge theory on a $1 \times 1$ lattice
with the periodic boundary condition
and it is equivalent to the SU($N$)
lattice gauge theory on the infinite lattice
in the planar limit.
We perform a Monte Carlo simulation of the model and find
that the model indeed allows
a different large $N$ limit, which
can be considered as the double scaling limit in matrix models.

The paper is organized as follows.
In section 2, we give the definition of the model and 
review some known results. We also explain the algorithm we use
in our simulation. 
In section 3, we study the planar limit of the model.
We show how the finite $N$ effects appear by measuring
Wilson loops. This gives us an important clue to a possible double 
scaling limit. 
In section 4, we present the data which show
the existence of the double scaling limit in the model.
Section 5 is devoted to conclusions and discussions. 

\section{The model}
\setcounter{equation}{0}
\label{sec:model}

The Eguchi-Kawai~(EK) model is defined by the following action \cite{EK}.
\begin{equation}
  \label{red-action}
  S_{EK}=-N\beta \sum_{\mu \ne \nu=1}^{D} 
%  Z_{\mu\nu} 
\mbox{tr} \{U_\mu U_\nu U^{\dag}_\mu U^{\dag}_\nu \},
\end{equation}
This model has the $U(1)^{D}$ symmetry.
\begin{equation}
U_\mu \to e^{i \theta_\mu} U_\mu.
\label{usym}
\end{equation}
In Ref. \cite{EK}, it has been shown that
if the $U(1)^{D}$ symmetry is not spontaneously broken,
the model is equivalent to the SU($N$) lattice gauge theory
on the infinite lattice in the large $N$ limit,
where the coupling constant $\beta$ in the action (\ref{red-action})
is kept fixed.
This limit is referred to as the planar limit, since in this limit
Feynman diagrams with planar topology dominate.

By equivalence, we mean that the expectation value of
the Wilson loop 
\begin{equation}
  \label{lat-W-loop}
  W(C)=\frac{1}{N} \mbox{tr} \left\{ U_{x,\alpha} U_{x+\hat{\alpha},\beta} 
    U_{x+\hat{\alpha}+\hat{\beta},\gamma} \cdots 
    U_{x-\hat{\lambda},\lambda} \right\},
\end{equation}
which is the product of link variables along a closed loop $C$,
calculated within the lattice gauge theory with the action
\begin{equation}
  \label{lat-action}
  S_{LGT}=- N \beta \sum_x \sum_{\mu \ne \nu=1}^D
  \mbox{tr} \{U_{x,\mu} U_{x+\hat{\mu},\nu}
  U^{\dag}_{x+\hat{\nu},\mu} U^{\dag}_{x,\nu} \},
\end{equation}
%where $U_{x,\mu}$'s are SU($N$) matrices and called link variables.
is equal to the expectation value of the
corresponding observable
\begin{equation}
  \label{W-loop}
   w(C)=\frac{1}{N} \mbox{tr} \left\{
    U_\alpha U_\beta U_\gamma \cdots U_\lambda \right\},
\end{equation}
calculated within the EK model with the action
(\ref{red-action}) in the planar limit.

It is found in Ref. \cite{BHN} that when the space-time dimension
$D$ is larger than two,
the U(1)$^{D}$ symmetry is 
spontaneously broken in the weak coupling region $\beta \gg 1 $.
This means that we cannot study the continuum limit of the
large $N$ gauge theory by the EK model when $D>2$,
since we have to send $\beta$ to infinity
when we take the continuum limit.
However, it is found that 
when $D$ is even,
one can twist the boundary condition of the EK
model so that the 
U(1)$^{D}$ symmetry is not broken in the weak coupling region,
while keeping the proof of the equivalence valid.
This is the twisted Eguchi-Kawai~(TEK) model \cite{TEK}
defined by the action
\begin{equation}
  \label{tred-action}
  S_{TEK}=-N\beta \sum_{\mu \ne \nu=1}^{D} 
  Z_{\mu\nu} 
\mbox{tr} \{U_\mu U_\nu U^{\dag}_\mu U^{\dag}_\nu \},
\end{equation}
where $Z_{\mu \nu}$ is called as the `twist', and defined as 
\begin{equation}
\label{twist}
  Z_{\mu\nu}=Z_{\nu\mu}^*= 
      \exp \left( 2 \pi i / L \right) \mbox{~~~~~~~for~~~} \mu<\nu,
\end{equation}
where $N=L^{D/2}$.
The observable that corresponds to the Wilson loop $W(C)$ is now given by
\begin{equation}
  \label{tW-loop}
%  W(C)\leftrightarrow 
w(C)=\frac{1}{N} 
\left( \prod_{\mu < \nu} Z_{\mu\nu}^{P_{\mu\nu}}  \right)
\mbox{tr} \left\{
    U_\alpha U_\beta U_\gamma \cdots U_\lambda \right\},
\end{equation}
where $P_{\mu\nu}$ 
is the number of plaquettes in the ($\mu\nu$) direction 
on the surface spanned by the Wilson loop $C$.
%For a general review of large $N$ reduced models,
%we refer the reader to Ref.\cite{Das}. 
The TEK model has been recently used for searching for nontrivial 
fixed points in the six-dimensional large $N$ gauge theory \cite{6D}.

In two dimensions, 
since the U(1)$^2$ symmetry is not spontaneously broken
even in the weak coupling region,
the EK model is equivalent to the lattice gauge theory
in the planer limit in the above sense.
The two-dimensional lattice gauge theory is solvable \cite{Gross-Witten}
in the planar limit
%large $N$ limit for fixed $\beta$ 
%in the action (\ref{lat-action})
and indeed the exact results obtained there
has been reproduced by the EK model in the planar limit \cite{EK}.
On the other hand, the TEK model can be considered 
in two dimensions as well, and 
its planar limit should be the same as that of the EK model.
Note, however, that the two models can show
different finite $N$ effects,
which is relevant to us,
since we are aiming at discovering a large $N$ limit
other than the planar limit.
Indeed we see that the approach to the planar limit from finite $N$
is quite different for the two models,
and the double scaling limit seems to exist only for the EK model.

Let us explain the algorithm we use for the Monte Carlo 
simulation of the EK and TEK models. 
Since the actions of the two models
(\ref{red-action}) and (\ref{tred-action}) 
are not linear in terms of 
link variables unlike that of the ordinary lattice gauge theory
(\ref{lat-action}),
we cannot apply the heat bath method \cite{heatbath} as it stands.
We therefore employ the technique proposed by Ref. \cite{FabriciusHaan},
and introduce the auxiliary field $Q_{\mu \nu}$ ($\mu<\nu$), 
which are general 
complex $N \times N$ matrices, with the following action.
\begin{eqnarray}
  \label{mod-action}
  S &=& N\beta \sum_{\mu<\nu} \tr Q_{\mu\nu}^{\dag} Q_{\mu\nu} \nonumber \\
    && -N\beta \sum_{\mu<\nu} \tr Q_{\mu\nu}^{\dag}
      \left( t_{\mu\nu} U_\mu U_\nu 
        + t_{\nu\mu} U_\nu U_\mu \right) \nonumber \\
    && -N\beta \sum_{\mu<\nu} \tr Q_{\mu\nu}
      \left( t_{\mu\nu}^* U_\mu^{\dag} U_\nu^{\dag}
        + t_{\nu\mu}^* U_\nu^{\dag} U_\mu^{\dag} \right),
\end{eqnarray}
where $t_{\mu\nu}$ is $1$ for the EK model and 
is $\sqrt{Z_{\mu \nu}}$ for the TEK model.
Integrating out the auxiliary field $Q_{\mu\nu}$, one reproduces
the original actions of the reduced models. 
The update of $Q_{\mu\nu}$ can be done easily 
by generating Gaussian variables.
Note that the action (\ref{mod-action})
is linear in terms of $U_\mu$, which means that
we can apply the heat bath method.
The update of $U_\mu$ is carried out by successively multiplying it by 
matrices each belonging to the $N(N-1)/2$ SU($2$) subgroups 
of SU($N$) \cite{CabibboMarinari}.
We define `one sweep' in this paper by
the update of all the elements of 
$Q_{\mu\nu}$'s and the update of all the $U_\mu$'s
by multiplying elements of every SU(2) subgroup once for each.
%For further technical details of the Monte Carlo algorithm,
%we refer the readers to Ref.\cite{FabriciusHaan}.

\section{Planar limit and the finite $N$ effects}
\setcounter{equation}{0}
\label{sec:planar}

In this section, 
we see how the EK model and 
the TEK model
at finite $N$ approach the known results in the planar limit 
as one increases $N$ with fixed $\beta$ in the action
(\ref{red-action}) and (\ref{tred-action}), respectively.

%As we mentioned in the previous section,
%the two reduced models are both equivalent to
%the ordinary lattice gauge theory 
%in the planar large $N$ limit.
%Two-dimensional SU($N$) lattice gauge theory
%is exactly solved in the planar limit \cite{Gross-Witten}

The observables calculated by Gross-Witten \cite{Gross-Witten}
for the two-dimensional SU($N$) lattice gauge theory in the planar limit
is the Wilson loop $W(C)$, where the contour $C$ is a rectangle.
Due to the discrete translational and rotational invariance
and the parity invariance of the lattice gauge theory,
the expectation value of Wilson loops is real and it
depends only on its shape but not on how the loop is placed on the
lattice nor on the orientation of the loop.
We therefore denote the expectation value of 
the $I \times J$ Wilson loop by $\langle W(I\times J) \rangle$.

We examine what we get for the corresponding observables 
(\ref{W-loop}) and (\ref{tW-loop})
in the reduced models with finite $N$.
%Before showing the results of Monte Carlo simulation,
%we remark the following.
Just as we mentioned above for the lattice gauge theory,
the expectation value of the observable in the EK model
that corresponds the Wilson loop is real and it depends only on its shape
even for finite $N$ due to the symmetries : 
$U_1 \rightarrow U_2$ and $U_2 \rightarrow U_1^{\dagger}$, 
which corresponds to the rotational symmetry, and
$U_1 \rightarrow U_2$ and $U_2 \rightarrow U_1$, which corresponds
to the parity invariance.
This is not the case, however, with the TEK model, 
in which the second one (`parity transformation') 
is not a symmetry.
Accordingly, the expectation value of
the observable in the finite $N$ TEK model
that corresponds to the Wilson loop is not real 
and it becomes complex conjugate when one flips the orientation of the
loop, though
the imaginary part should vanish in the planar limit,
where it should reproduce the Wilson loop in the lattice gauge theory.
%Since the first one (`rotational symmetry') is a symmetry of the TEK model,
%the above quantity does not depend on how the loop is placed on the lattice.
When we define $\langle w(C) \rangle$ in reduced models 
which corresponds to $\langle W(C) \rangle$ in the 
lattice gauge theory,
the real part is implicitly taken in the case of the TEK model.
When we measure $\langle w(C) \rangle$ in either model,
we average over the transformation of the Wilson loop
by rotation and parity in order to increase the statistics.

The internal energy of the system can be given 
by the expectation value of the $1 \times 1$ Wilson loop
\begin{equation}
  \label{1pq}
  E(\beta) =  \langle W(1 \times 1) \rangle ,
\end{equation}
and the analytical result in the planar limit is obtained as 
\cite{Gross-Witten}
\begin{equation}
  \label{Internal-Energy}
  E(\beta)= \left\{
    \begin{array}{ll}
      \beta & \;\; \mbox{for}\;\;\; \beta \le\frac{1}{2} \\
      1-\frac{1}{4\beta} &  \;\; \mbox{for}\;\;\; \beta\ge\frac{1}{2}.
    \end{array}
    \right.
\end{equation}
As one sees from this result, the second derivative of the
internal energy is discontinuous at $\beta=1/2$,
which means that the system undergoes a 
third-order phase transition at $\beta=1/2$.
%In the following,
%we restrict ourselves to $\beta$ larger than the critical point,
%{\it i.e.} $\beta > 1/2$,
%since we have to take the continuum limit by sending $\beta$ to
%infinity, as we see later.
%when we discuss the scaling behavior.

\begin{figure}[p]
\begin{center}
\leavevmode\psfig{figure=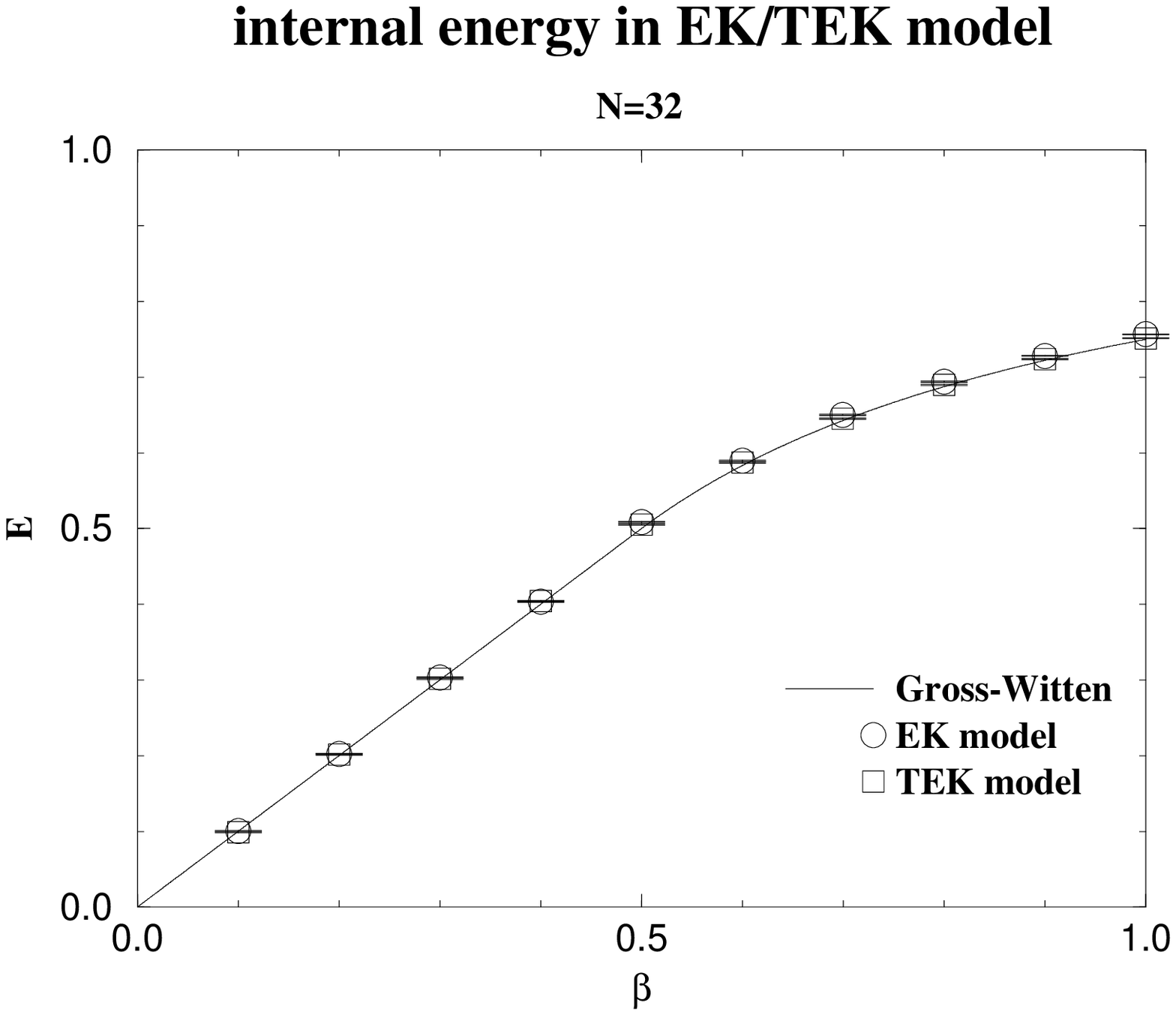,width=16cm,height=12.8cm}
\end{center}
\caption[internal energy]{The internal energy
is plotted against $\beta$ for the EK model (circles)
and for the TEK model (squares) for $N=32$. 
The solid line represents the theoretical result in the
planar limit ($N\rightarrow \infty$ with 
fixed $\beta$) obtained by Gross-Witten.}
\label{test}
\end{figure}

In Fig. \ref{test}, 
we plot the internal energy against $\beta$
for the two models for $N=32$.
We can see that the data agree nicely
to the theoretical result in the planar limit 
(\ref{Internal-Energy}), which is represented
by the solid line.
In Ref. \cite{TEK}, the results for $N=20$ 
are given, which shows that the data for the EK model are slightly 
larger than the planar result, while the data for the TEK model are
in complete agreement with it.
This tendency has been observed in our simulation as well with smaller $N$. 
Thus as far as the internal energy is concerned, the convergence to
the planar limit is better for the TEK model than for the EK model.

\begin{figure}[p]
\begin{center}
\leavevmode\psfig{figure=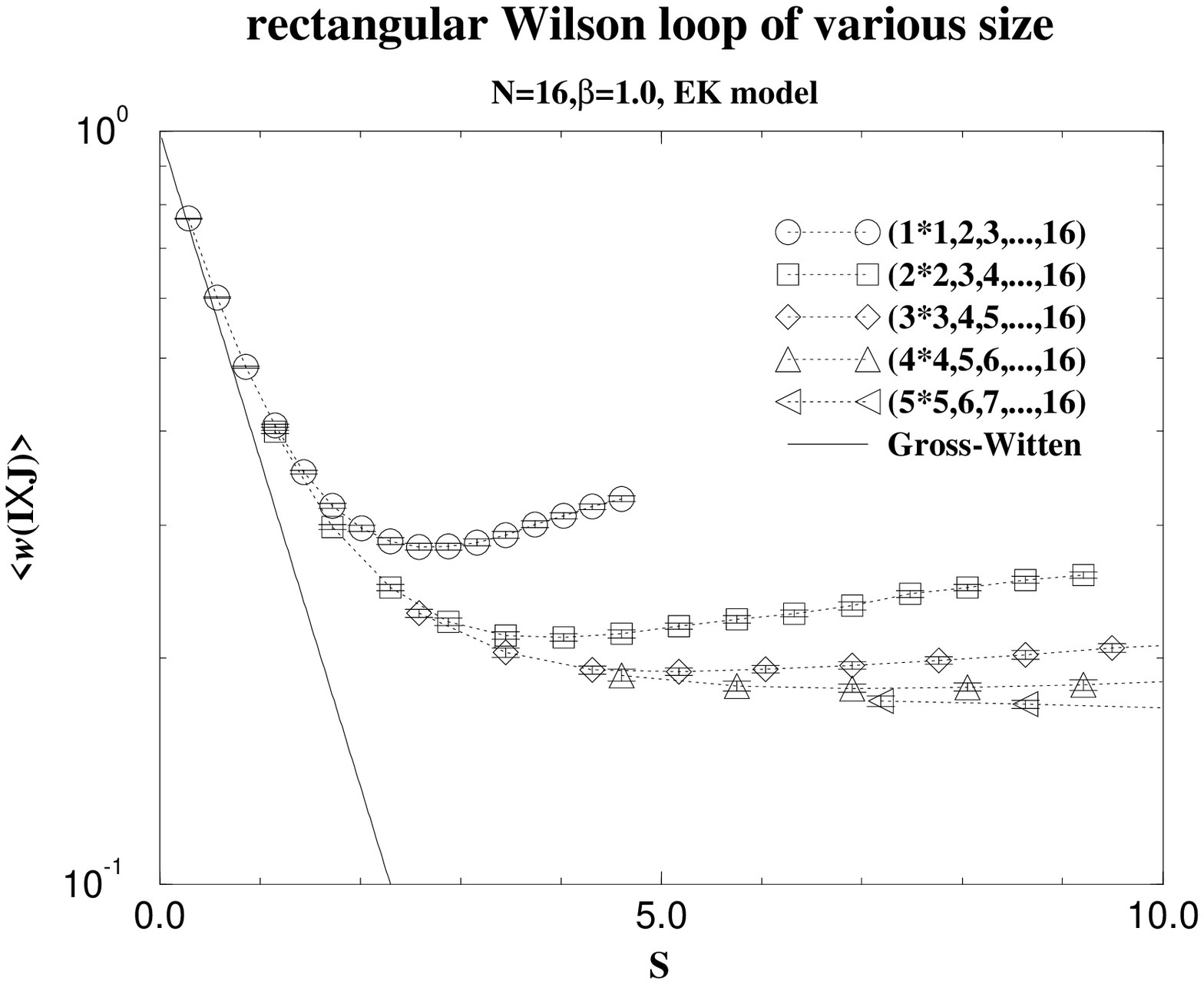,width=16cm,height=12.8cm}
\end{center}
\caption[Wilson loop amplitude as a function of the area for the EK model]
{The expectation value of $w(I\times J)$ in the EK model 
($N=16$, $\beta =1.0$)
which corresponds to $I\times J$ Wilson loops
is plotted against the physical area $S=a^2 IJ$.
Each symbol represents the data for $I=1$ (circles), $I=2$ (squares),
$I=3$ (diamonds), $I=4$ (triangles) and 
$I=5$ (tilted triangles) with $J\ge I$.
The straight solid line represents the planar result by Gross-Witten.}
%The dotted lines are drawn for each symbol to guide the eye.
\label{nekwlp1}
\end{figure}

Let us turn to the expectation value of 
rectangular Wilson loops of various size.
The planar result is given by
\begin{equation}
  \label{WilsonLoop}
  \langle W(I \times J) \rangle 
= \exp \left( -\kappa(\beta) I J \right),
\end{equation}
where
\begin{equation}
  \label{tension}
  \kappa(\beta) =\left\{
    \begin{array}{ll}
      -\log\beta & \;\;\;\mbox{for}\;\;\; \beta\le\frac{1}{2} \\
      -\log\left(1-\frac{1}{4\beta}\right) &
      \;\;\;\mbox{for}\;\;\; \beta\ge\frac{1}{2}.
    \end{array}
    \right.
\end{equation}
For $I=J=1$, the above result reduces to the one for the internal energy
(\ref{Internal-Energy}).
This result shows that the rectangular Wilson loops obey the area law
exactly for all $\beta$.
From this result, one can figure out how to fine-tune the coupling
constant $\beta$ as a function of the lattice spacing $a$ 
when one takes the continuum
limit $a \rightarrow 0$.
Since the physical area is given by $S=a^2 IJ $, we have to fine-tune
$\beta$ so that $\kappa(\beta)/a^2$ is kept fixed.
We therefore take 
\begin{equation}
\label{ptension}
a=\sqrt{\kappa(\beta)}
\end{equation}
in the following.
Since $\kappa(\beta)$ is given by eq. (\ref{tension}),
we have to send $\beta$ to infinity as we take the $a \rightarrow 0$
limit.

In Fig. \ref{nekwlp1}, we plot the expectation value of
$I \times J$ Wilson loops in the EK model
against the physical area $S=a^2 IJ$
for $N=16$ and $\beta=1.0$, where $I$ and $J$ run from $1$ to $16$.
Although the planar result depends only on the area of the Wilson 
loop, but not on its shape,
we see that the results for finite $N$ depend on the shape as well.
For example, we see that the result for the $1 \times 9$ Wilson loop
is farther from the planar limit
than that for the $3 \times 3$ Wilson loop.
In general, one sees that the finite $N$ effect for Wilson loops
with the same area becomes larger
as it becomes longer in one direction.
This means that we have to specify not only the area
but also the shape of the Wilson loop when
we discuss the scaling behavior in the double scaling limit,
as well as when we discuss the finite $N$ effects.
%In the following, we restrict ourselves to square Wilson loops
%namely with $I=J$ for simplicity.

\begin{figure}[p]
\begin{center}
\leavevmode\psfig{figure=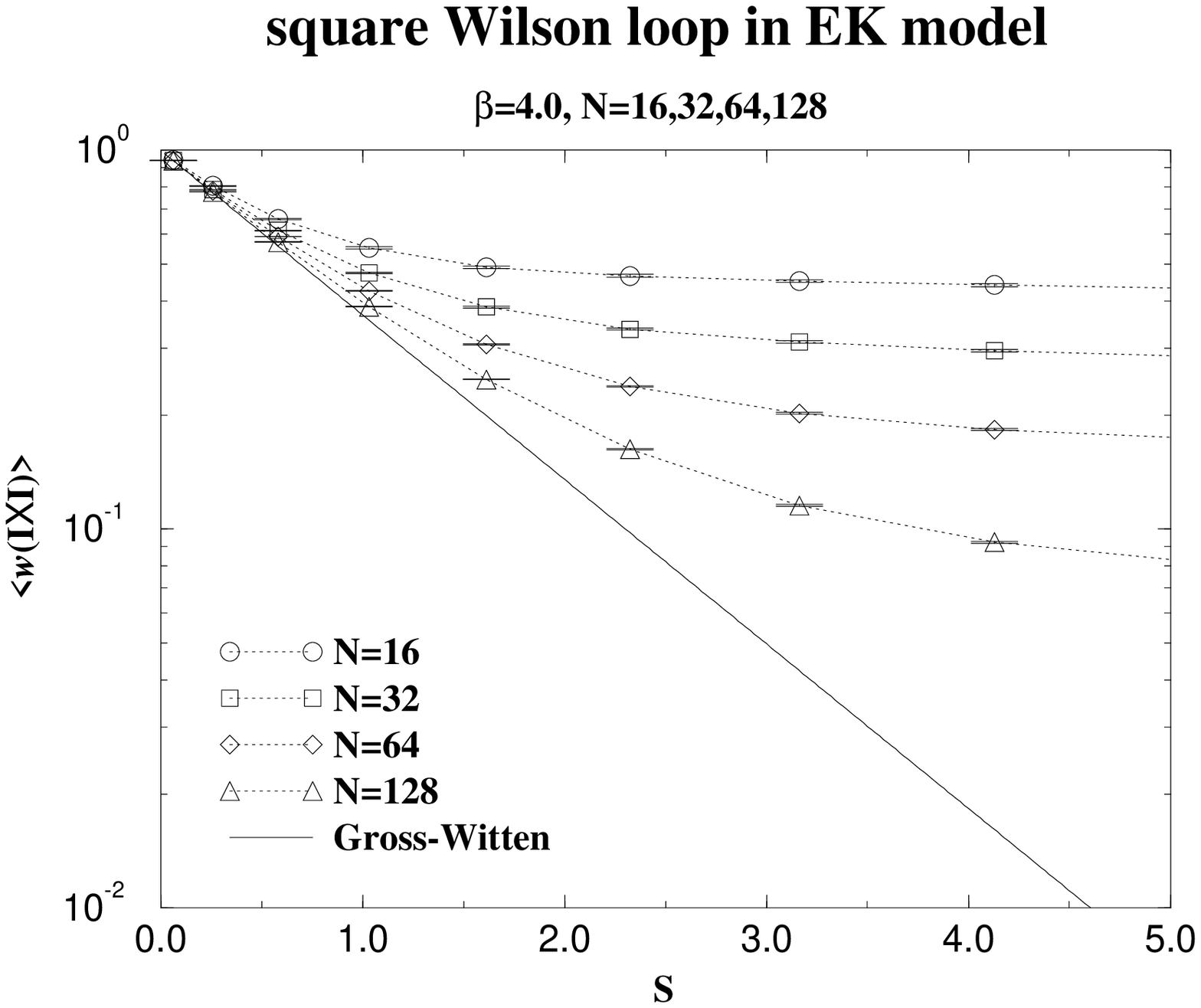,width=16cm,height=12.8cm}
\end{center}
\caption[Wilson loop amplitude as a function of the area for the EK model]{
The expectation value of $w(I\times I)$ in the EK model 
for $\beta =4.0$ is plotted against the physical area $S=(a I)^2$.
Each symbol represents the data for $N=16$ (circles), $N=32$ (squares),
$N=64$ (diamonds) and $N=128$ (triangles).
The straight solid line represents the planar result by Gross-Witten.}
\label{nekwlp2}
\end{figure}

\begin{figure}[p]
\begin{center}
\leavevmode\psfig{figure=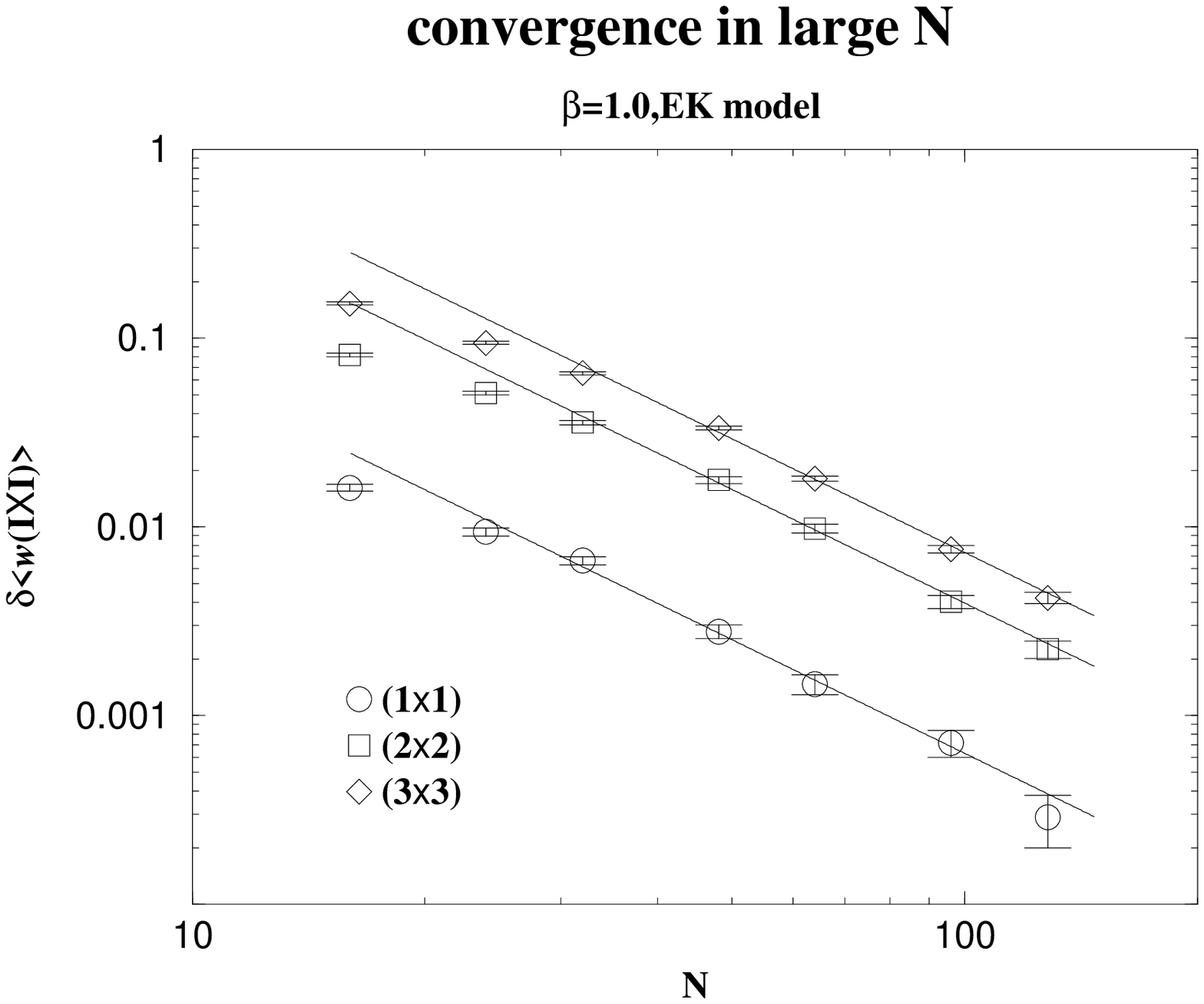,width=16cm,height=12.8cm}
\end{center}
\caption[Wilson loop amplitude as a function of $N$ for the EK model]{
The deviation of 
the expectation value of $w(I\times I)$ in the EK model 
from the planar result 
is plotted against $N$ for $I=1$ (circles), $I=2$ (squares) and
$I=3$ (diamonds) at $\beta =1.0$.
The solid straight line represents a fit to $const./N^2$,
which is the dominant subleading term in the large $N$ limit.}
\label{wlpbeh}
\end{figure}

In Fig. \ref{nekwlp2}
% and Fig.\ref{wlpbeh}, 
we plot the expectation value of square Wilson loops
in the EK model
against the physical area $S=(aI)^2$
for $\beta=4.0$ with $N=16,32,64$ and $128$.
One can see that the data points approach the planar result
from above monotonously.
In Fig. \ref{wlpbeh} we plot the deviation of the data 
for fixed Wilson loops from the planar result
as a function of $N$ for $\beta = 1.0$.
One can see a clear $1/N^2$ behavior, which means that 
the deviation is dominated by the subleading term in the large $N$ limit.
For $\beta = 4.0$, the deviation decreases slower than $1/N^2$
up to $N=128$, which means that the contribution from the higher order
terms is still comparable with the subleading term.
The double scaling limit we consider in the next section corresponds
to taking the large $N$ limit together with the $\beta \rightarrow
\infty$ limit so that all the terms in the $1/N^2$ expansion
contribute.

\begin{figure}[p]
\begin{center}
\leavevmode\psfig{figure=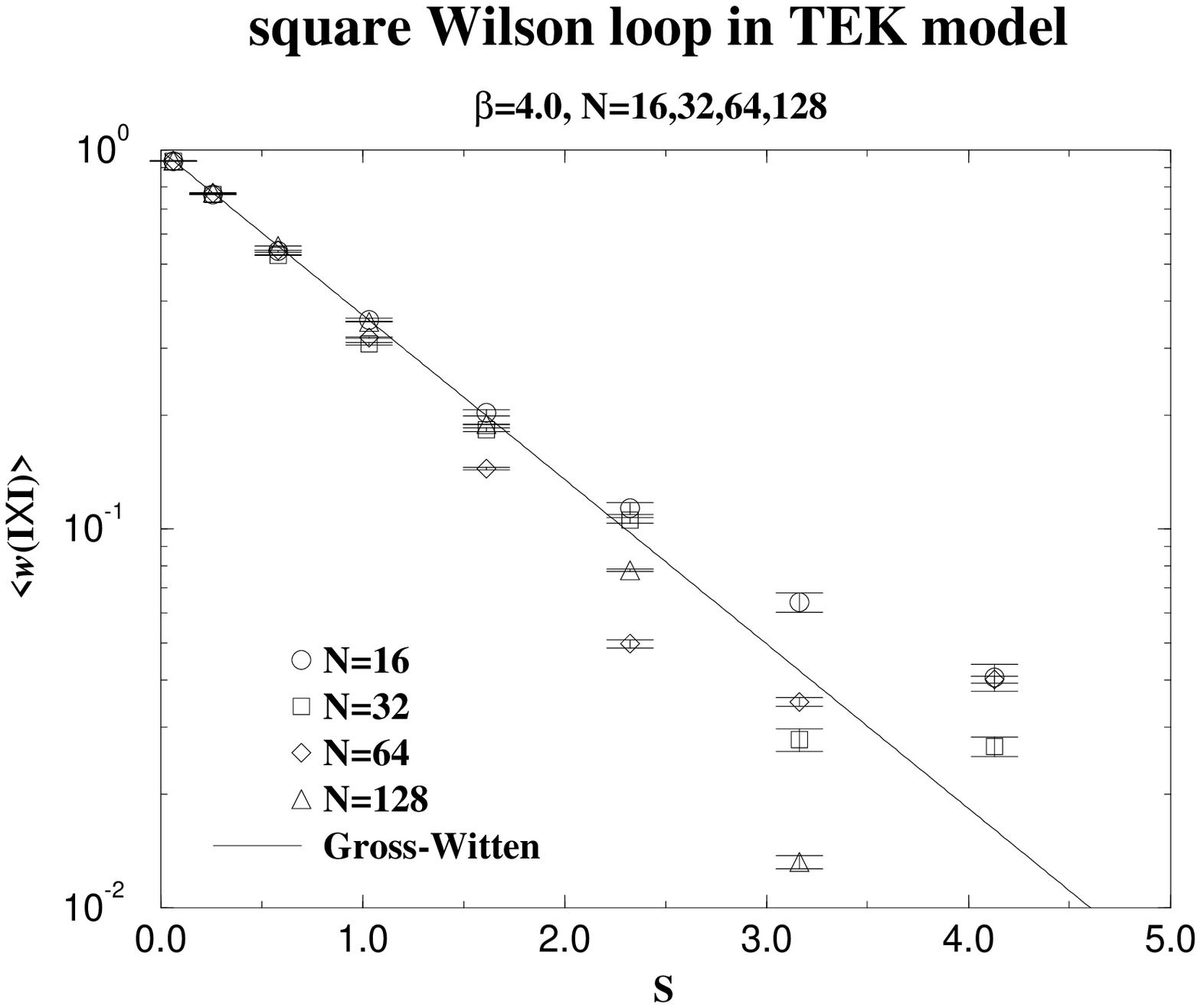,width=16cm,height=12.8cm}
\end{center}
\caption[Wilson loops as a function of area for the TEK model]{
The expectation value of $w(I\times I)$ in the TEK model 
for $\beta =4.0$ is plotted against the physical area $S=(a I)^2$.
Each symbol represents the data for $N=16$ (circles), $N=32$ (squares),
$N=64$ (diamonds) and $N=128$ (triangles).
The straight solid line represents the planar result by Gross-Witten.}
\label{tekwlp}
\end{figure}

The finite $N$ effects for the TEK model appear in quite a 
different manner.
Fig. \ref{tekwlp} shows the results for the TEK model
for the same set of values of $N$ and with the same $\beta$
as in Fig. \ref{nekwlp2}.
The data point for the largest $S$ for $N=128$ is omitted,
since the result turned out to be negative within the error bar.
We have checked the thermalization by using hot starts and
cold starts. The number of sweeps needed to achieve the
thermalization is less than 1000 sweeps for $N=128$.
The errorbars are estimated using the standard 
jackknife method. The autocorrelation time is approximately
200 sweeps for $N=128$.
One sees from the results for fixed area with $S<2.5$
that the data do not approach the planar
result monotonously, but rather show some overshooting behavior.
For larger $S$, the tendency of convergence 
is not yet seen up to $N=128$.
Thus, although it might be that 
the convergence to the planar limit
is somewhat accelerated in the TEK model as is seen 
from the results 
for smaller Wilson loops,
we find that the finite $N$ effects in the TEK model are much more 
complicated than in the EK model.

\begin{figure}[htb]
\begin{center}
\leavevmode\psfig{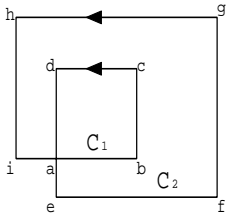}
\end{center}
\caption[]{A typical Wilson loop with a self-interaction}
\label{dbllp}
\end{figure}

Another interesting quantity known in the planar limit
is the expectation value of Wilson loops with self-intersections 
\cite{KK}. 
In Fig. \ref{dbllp}, we illustrate a typical Wilson loop we consider.
It is given by the trace of the product of link variables along
the loop $C={abcdaefghia}$.
We split the original loop $C$ into two loops at the 
self-intersection point and define
$C_1={abcda}$ and $C_2={aefghia}$.

\begin{figure}[p]
\begin{center}
\leavevmode\psfig{figure=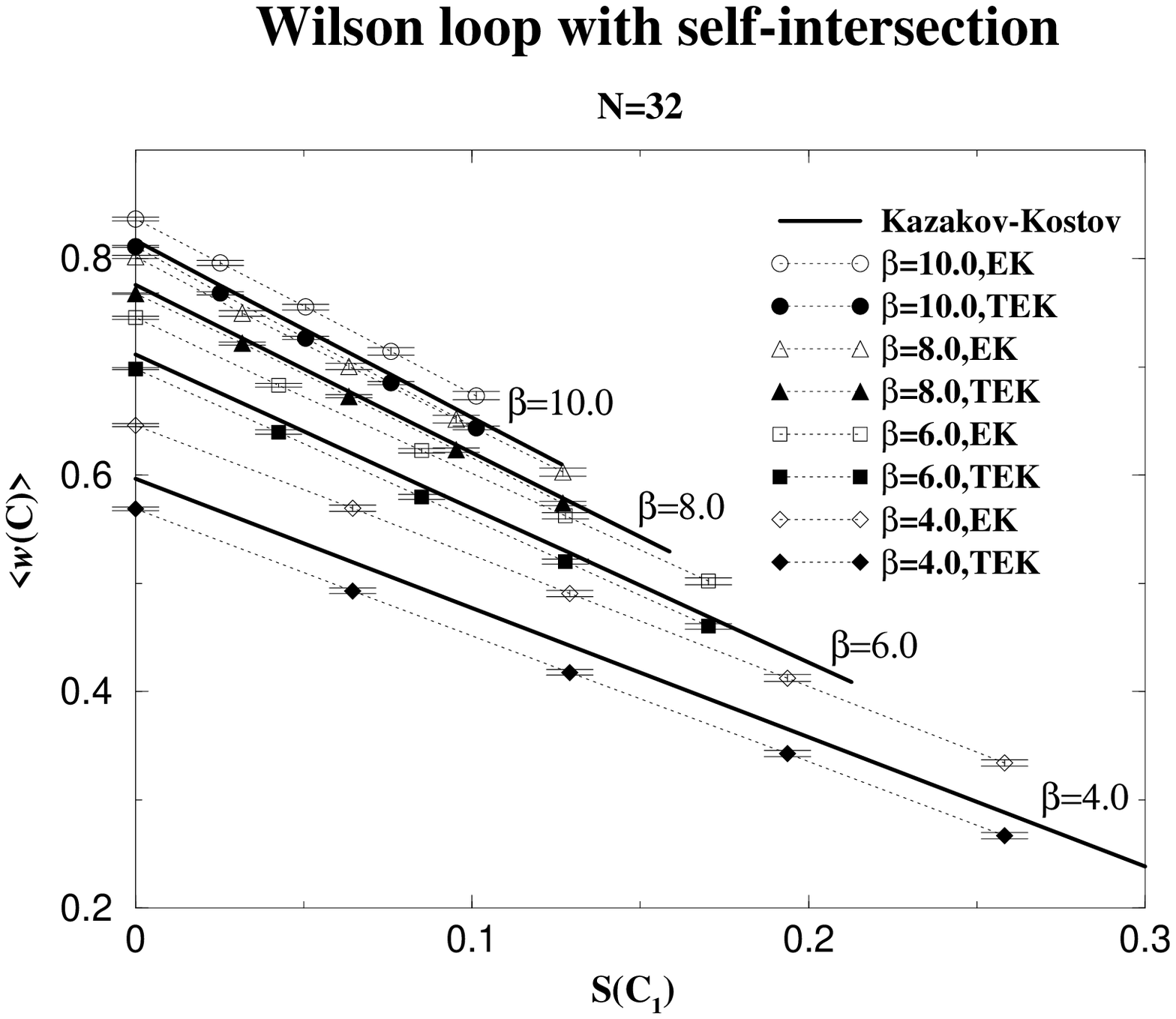,width=16cm,height=12.8cm}
\end{center}
\caption[double loop amplitude for $N=32$]{
The expectation value of $w(C)$ in the EK (open symbols) and the 
TEK model (filled symbols) for $N=32$
with $\beta=4.0$ (diamonds), $6.0$ (squares), $8.0$ (triangles) and
$10.0$ (circles)
are plotted against the physical area $S(C_1)=a^2  L L_1$.
The straight heavy solid lines represent the continuum results in the 
planar limit by Kazakov-Kostov.}
\label{dlp32}
\end{figure}

\begin{figure}[p]
\begin{center}
\leavevmode\psfig{figure=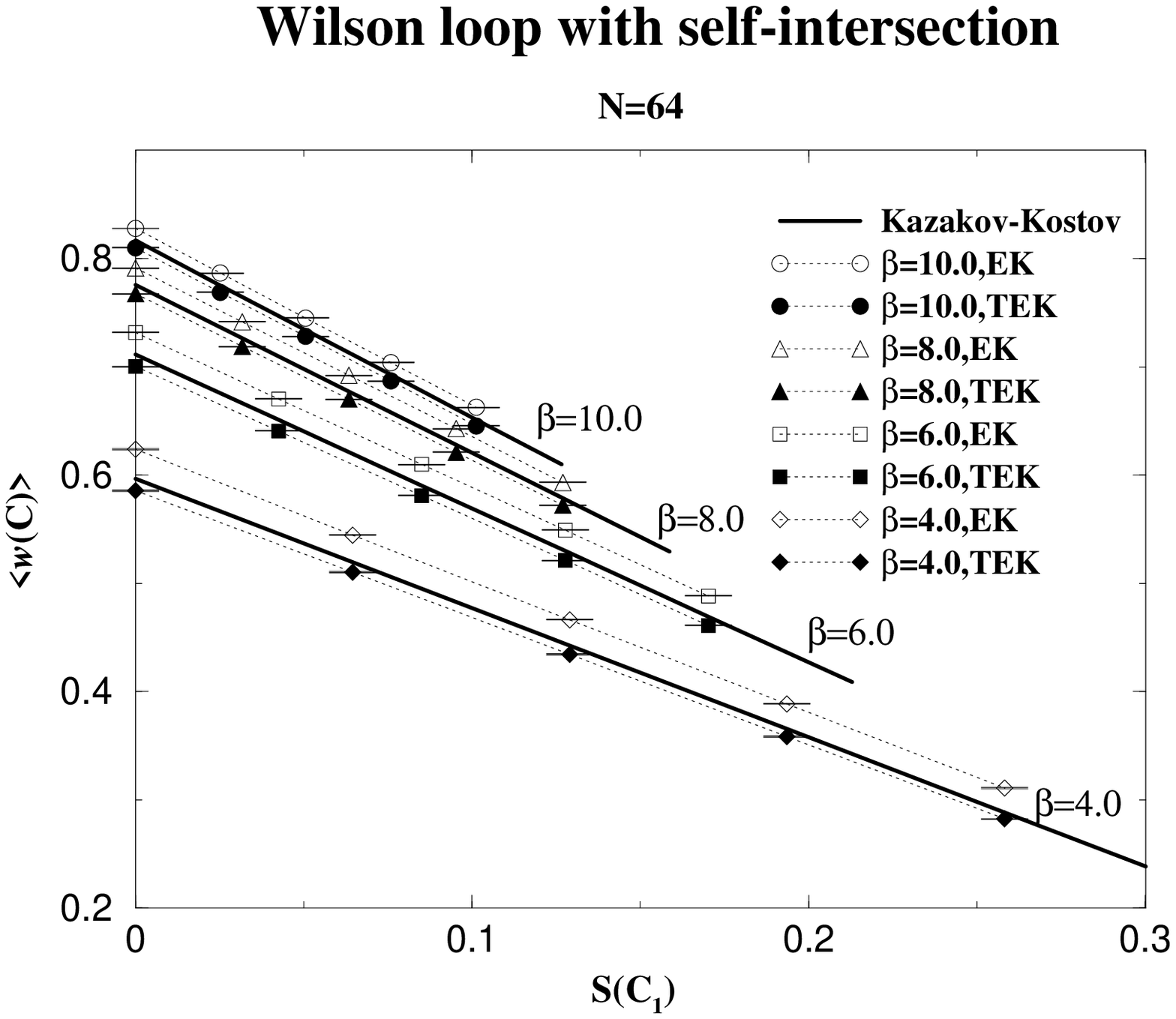,width=16cm,height=12.8cm}
\end{center}
\caption[double loop amplitude for $N=64$]{
The expectation value of $w(C)$ in the EK (open symbols) and the 
TEK model (filled symbols) for $N=64$
with $\beta=4.0$ (diamonds), $6.0$ (squares), $8.0$ (triangles) and
$10.0$ (circles)
are plotted against the physical area $S(C_1)=a^2 L  L_1$.
The straight heavy solid lines represent the continuum results in the 
planar limit by Kazakov-Kostov.}
\label{dlp64}
\end{figure}

The expectation value of such a Wilson loop has been
calculated \cite{KK} 
by solving the Makeenko-Migdal equation \cite{MM} 
in the continuum and the result is given by
\begin{equation}
  \label{KK}
  \langle W(C) \rangle = 
    (1- 2 K S(C_1)) \exp \left[ - K\{S(C_1)+S(C_2)\}\right]  .
\end{equation}
$S(C)$ is the physical area of the region surrounded by the loop $C$.
When $S(C_1)=0$, the above expression reduces to the one for
the Wilson loop without self-interactions, which obeys the area law.
$K$ is the physical string tension, which is taken to be $1$
in this paper, since we have fixed the scale by eq. (\ref{ptension}).
When $S(C_1)>0$, the factor $(1- 2 K S(C_1))$ gives a nontrivial 
correction.
We measure the observable 
\begin{equation}
  w(C)=\frac{1}{N} \mbox{tr} \left(
    U_1^L U_2^{L_1} U_1^{\dag L} U_2^{\dag L_1}
    U_1^L U_2^{L_2} U_1^{\dag L} U_2^{\dag L_2} \right), \;\;\;L_1 \le L_2,
\end{equation}
in the EK model and the corresponding one
in the TEK model, where we multiply the above expression
by $Z_{12}^{L(L_1+L_2)}$ according to the definition (\ref{tW-loop}).
Comparison with the continuum result in the planar limit
(\ref{KK}) can be made by putting
$S(C_1)=a^2 L L_1$ and $S(C_2)=a^2 L L_2$.
In order to extract the nontrivial factor $(1-2 K S(C_1))$, where $K=1$,
we measure the expectation value $\langle w(C) \rangle $
with fixed $S(C_1)+S(C_2)$ and plot the result against $S(C_1)$.
Specifically, we take $L=1$ and consider $L_1$ and $L_2$ with
$L_1 + L_2 = 8$ and $L_1=0,1,2,3,4$.
The observable for $L_1=0$ is nothing but the $(L\times L_2)$ Wilson loop.
Fig. \ref{dlp32} and Fig. \ref{dlp64} 
shows the results for $N=32$ and $N=64$ respectively.
The planar results by Kazakov-Kostov are shown by the straight 
heavy solid lines.
One sees that the data are on the lines
parallel to the planar results, and the lines are
slightly above the planar results for the EK model,
and slightly below the planar results for the TEK model
\footnote{Strictly speaking, $L$, $L_1$ and $L_2$ should be large
enough when we compare the data with the continuum result.
The data show, however, that the finite lattice spacing effect is
absent accidentally for the observable considered, 
as in the case with the simple rectangular Wilson loops, which obey
the area law exactly even for the $(1 \times 1)$ loop.}.
The discrepancy is smaller for $N=64$ than for $N=32$.


This shows that the finite $N$ effect comes mainly from the area law part,
and the correction due to the factor $(1- 2K S(C_1))$ in (\ref{KK}) 
does not show any significant finite $N$ effect.
%In view of this, we do not further examine the Wilson loops with
%self-intersections when we consider the double scaling limit in the next 
%section.
We also note that the total physical area 
$S(C_1)+S(C_2)$ is 0.516 for $\beta=4.0$
and the tendency of convergence to the planar limit is 
consistent with that observed in Fig. \ref{nekwlp2} and Fig. \ref{tekwlp}.

\section{Existence of double scaling limit in EK model}
\setcounter{equation}{0}
\label{sec:dsl}

In this section, we consider the EK model
not as a reduced model which reproduce the lattice gauge theory 
in the planar limit, but as a matrix model which defines a string theory.
We therefore examine whether we can take a large $N$ limit
other than the planar limit discussed in the previous section.

We consider the EK model as a toy model of the IIB matrix model.
Just as in Ref. \cite{IKKT,FKKT}, we make the T-duality transformation,
when we interpret the EK model as a string theory.
Upon the T-duality transformation, the compactification radius of the
torus to which the space time is compactified
is inversed and the winding mode and the momentum mode are exchanged.
Since the EK model can be considered to be
defined on a unit cell of size $a$ 
with the periodic boundary condition,
it can be considered, after the T-duality transformation,
as a string theory in the two-dimensional space time
compactified on a torus of size $1/a$.

The observables we consider in this section is the Wilson loops
\beq
\mbox{tr} \left\{ U_\alpha U_\beta U_\gamma \cdots U_\lambda \right\},
\label{stringf}
\eeq
where the suffix $\alpha,\beta,\cdots,\lambda$ runs over
$\pm 1$,$\pm 2$ and $U_{-\alpha}$ ($\alpha > 0$) is defined by
$U_{-\alpha} = U_{\alpha}^{\dagger}$.
The winding number of the Wilson loop is given by
$n_\mu=\hat{\alpha}_\mu+\hat{\beta}_\mu+\cdots + \hat{\lambda}_\mu$,
where $\hat{\alpha}_\mu$ denotes a unit vector in the $\alpha$ direction.
Note that the observables (\ref{W-loop})
considered in the previous section can be viewed as
the Wilson loops in the EK model with $n_\mu =0$,
while in this section we consider those with
$n_\mu \ne 0 $ as well.
After the T-duality transformation, 
the winding of the Wilson loops represents
the momentum distribution of the string in the two-dimensional space time.
The total physical momentum carried by the string
which corresponds to the Wilson loop (\ref{stringf}) is given by
$P_\mu = n_\mu a$.
Under the $U(1)^2$ transformation (\ref{usym}),
the Wilson loop (\ref{stringf}) is multiplied by
$\exp (i n_\mu \theta_\mu) $.
Since the symmetry is not spontaneously broken in the present model,
the one-point function of Wilson loops in the winding sector
should vanish. 
Similarly, the $n$-point function should vanish in general, 
unless the sum of the winding number of the Wilson loops is zero.
After the T-duality transformation,
the above selection rule 
gives nothing but the momentum conservation in the 
two-dimensional space time.
This corresponds to the fact that
%is natural since 
the U(1)$^2$ symmetry is nothing but the 
translational invariance when we identify the phase of the eigenvalues of
$U_\mu$ as the two-dimensional space-time coordinates.

In what follows, we consider 
\beqa
O^{(n)}_{\alpha\beta}(L) &=& \tr(U_\alpha ^L U_\beta ^L 
                U_{-\alpha} ^L U_{-\beta} ^L  ) \n
O^{(w)}_{\alpha}(L) &=& \tr(U_\alpha ^L),
\label{typicalO}
\eeqa
which give typical Wilson loops in the non-winding sector
and the winding sector respectively.
Due to the trace, the independent operators in the non-winding sector
are actually given by $ O^{(n)}_{12}(L)$ and $O^{(n)}_{21}(L)$,
which are complex conjugate to each other.
When we take the $a\rightarrow 0$ limit,
we have to send $L$ to the infinity by
fixing $aL$.
Note that in (\ref{stringf}) and (\ref{typicalO}),
we do not put $1/N$ in front of the trace as we did in (\ref{W-loop}).
This is because with this convention, the power counting of $N$ 
for correlation functions matches with the one conventional in the
old-fashioned matrix model \cite{DSL}.

We do not consider the TEK model in what follows, since 
the interpretation as a string theory is not as obvious as in the EK model.
Indeed, as is seen in Fig. \ref{tekwlp},
the TEK model does not show any systematic finite $N$ effects, 
while the EK model does.
It is therefore natural to consider that the TEK model does not allow any
sensible double scaling limit.
%the EK model shows a systematic finite $N$ effects, 
%while the TEK model does not.

\begin{figure}[p]
\begin{center}
\leavevmode\psfig{figure=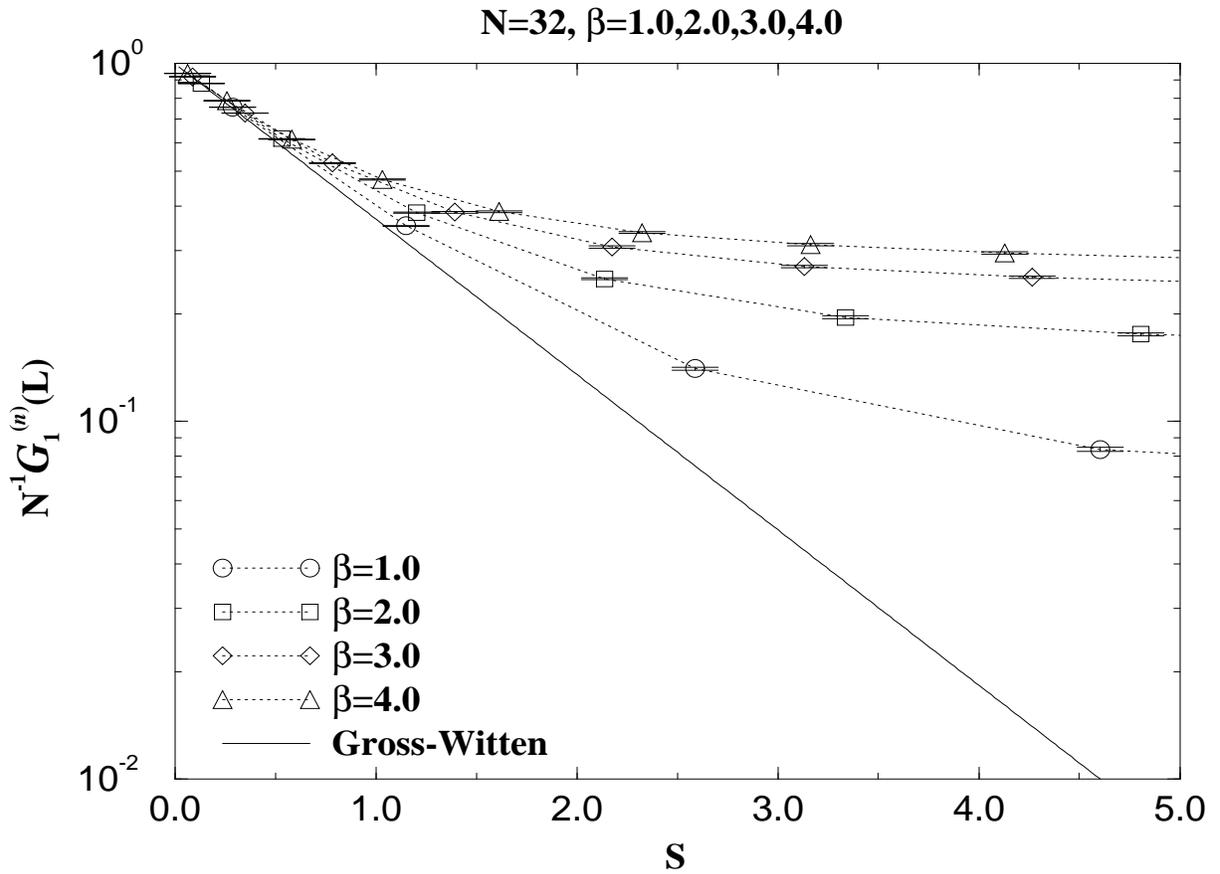,width=16cm,height=12.8cm}
\end{center}
\caption[Wilson loop amplitude varying $\beta$]{
The one-point function of the Wilson loops
in the non-winding sector $N^{-1}G_1^{(n)}(L)$ 
in the EK model is plotted
against $S=(aL)^2$ for $N=32$ 
with $\beta=1.0$ (circles), $2.0$ (squares),
$3.0$ (diamonds) and $4.0$ (triangles).
The straight solid line represents the planar result by Gross-Witten.}
\label{nekwlp3}
\end{figure}

\begin{figure}[p]
\begin{center}
\leavevmode\psfig{figure=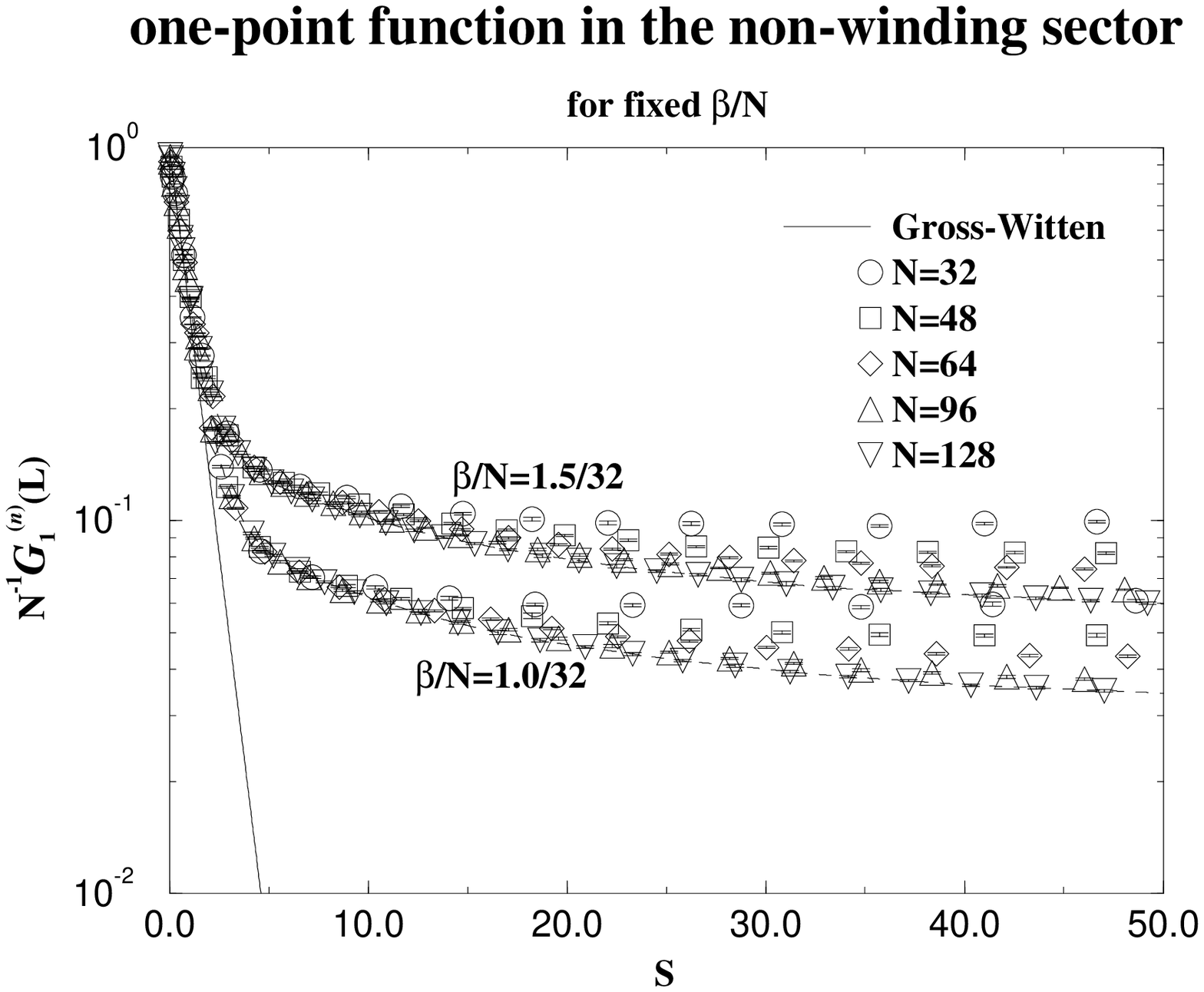,width=16cm,height=12.8cm}
\end{center}
\caption[Wilson loop amplitude in the double scaling limit]{
The one-point function of the Wilson loops
in the non-winding sector $N^{-1}G_1^{(n)}(L)$ 
in the EK model is plotted
against $S=(aL)^2$ for 
the sets of $\beta$ and $N$ with $\beta/N$ fixed to 
$1.0/32$ and $1.5/32$.
Each symbol represents the data for $N=32$ (circles), $N=48$ (squares),
$N=64$ (diamonds), $N=96$ (triangles) and $N=128$ (inverse triangles).
The straight solid line represents the planar result by Gross-Witten.
The dashed lines, which connect the data for $N=128$,
are drawn to guide the eye.
}
\label{dblwlp}
\end{figure}

Let us first examine the one-point function
of Wilson loops in the non-winding sector, namely
$G_1^{(n)}(L) = \langle O^{(n)}_{12}(L) \rangle $.
The observable $\langle w(L \times L) \rangle $ 
considered in the previous section
is nothing but $N^{-1} G_1^{(n)}(L)$.
%As is mentioned in the previous section,
%$\langle O^{(n)}_{12}(L) \rangle$ is equal to 
%$\langle O^{(n)}_{21}(L) \rangle$, due to the symmetry of the model,
%and therefore they are both real.
%In the measurement, we take the average of the two quantities
%in order to increase the statistics.
We have seen in Fig. \ref{nekwlp2}
how the data for $N^{-1}  G_1^{(n)}(L)$ 
approach the planar limit when we increase $N$ with fixed $\beta$.
In Fig. \ref{nekwlp3}, we plot 
$N^{-1} G_1^{(n)}(L)$
against $S=(a L)^2$ for fixed $N$ with various values of $\beta$.
One can see that the effect of increasing $\beta$ is quite similar to
the finite $N$ effect.
Thus it is natural to expect that we may be able to balance these two effects 
by taking the limit $\beta \to \infty$ together with the $N \to \infty$
limit.
Indeed we find that 
$N^{-1}  G_1^{(n)}(L)$ scales in the large $N$ limit
by fixing $\beta/N$.
In Fig. \ref{dblwlp} we show the data
for $\beta/N =1.0/32$ and $\beta/N =1.5/32$.
$N$ is taken to be $32,48,64,96$ and $128$.
We see a clear scaling behavior.
Measurements have been done every 10 sweeps and
we average over 2000 configurations
for $N=32\sim96$, and
over 1000 configurations for $N=128$.
All the data in this section have been obtained in this way,
unless mentioned.

The finite $N$ effect is seen as a deviation from the scaling
behavior for large $S$.
The deviation shows up at $L\simeq N/8$.
Since $a\propto 1/\sqrt{\beta} \propto 1/\sqrt{N}$ for large $N$,
the scaling region enlarges as $\sqrt{N}$.
The scaling function for fixed $\beta/N$ approaches 
the planar limit as $\beta/N$ becomes smaller.
%, which is to be expected.

From this observation, it is natural to expect that
the double scaling limit of the EK model can be taken by
fixing $\beta/N$, which corresponds to 
the string coupling constant $g_{str}$.
The planar limit corresponds to $g_{str}=0$.

Let us next consider Wilson loops in the winding sector.
Since the one-point function vanishes as is explained above,
we consider the two-point function
\begin{equation}
  \label{opn2pt}
  G_2^{(w)}(L) = 
\langle O_\mu^{(w)}(L) O_{-\mu}^{(w)}(L) \rangle . 
\end{equation}
Note that the right hand side does not depend on $\mu$, 
due to the rotational invariance.
In the measurement, we average over $\mu=1,2$ to increase the statistics.
This quantity is naively ${\cal O}(N^2)$, but
due to the factorization, its leading term
is given by
$\langle O_\mu^{(w)}(L)\rangle   \langle 
 O_{-\mu}^{(w)}(L) \rangle $,
which actually vanishes due to the U(1)$^2$ symmetry.
The order of the quantity is therefore ${\cal O}(1)$.
We can regard the ${\cal O}(1)$ quantity as
the connected two-point function of the Wilson loops in the 
winding sector.

\begin{figure}[p]
\begin{center}
\leavevmode\psfig{figure=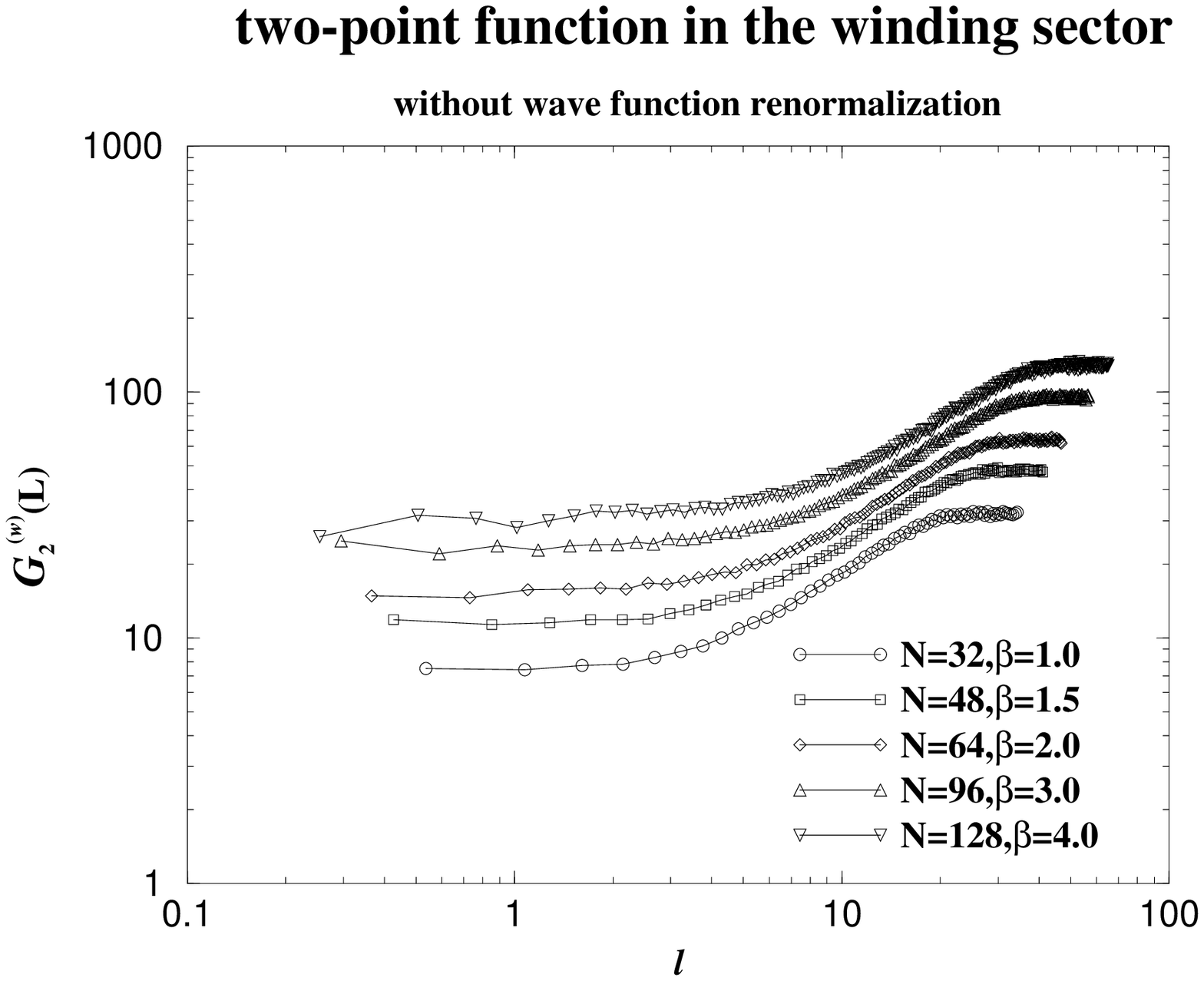,width=16cm,height=12.8cm}
\end{center}
\caption[open loop 2-point function without the wave function renormalization]{
The two-point function in the winding sector 
without the wave function renormalization
$G_2^{(w)}(L)$ 
is plotted against $l=aL$
for sets of $\beta$ and $N$ with $\beta/N=1.0/32$.
Each symbol represents the data for $N=32$ (circles), $N=48$ (squares),
$N=64$ (diamonds), $N=96$ (triangles) and $N=128$ (inverse triangles).}
\label{opn2ptnr}
\end{figure}

\begin{figure}[p]
\begin{center}
\leavevmode\psfig{figure=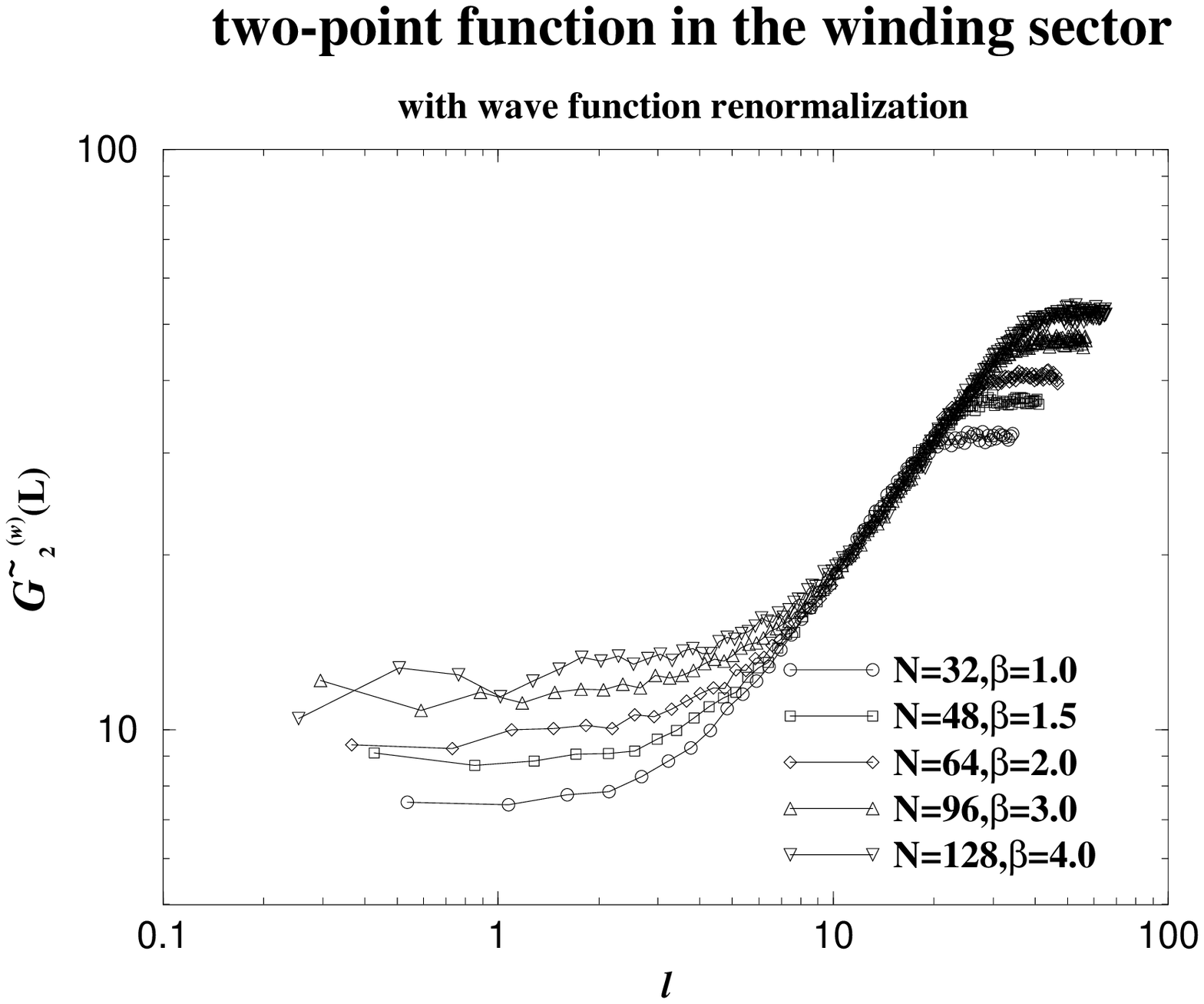,width=16cm,height=12.8cm}
\end{center}
\caption[open loop 2-point function with the wave function renormalization]{
The two-point function in the winding sector 
with the wave function renormalization
$\widetilde{G}_2^{(w)}(L)$ is plotted against $l=aL$
for sets of $\beta$ and $N$ with $\beta/N=1.0/32$.
Each symbol represents the data for $N=32$ (circles), $N=48$ (squares),
$N=64$ (diamonds), $N=96$ (triangles) and $N=128$ (inverse triangles).}
\label{opn2ptr}
\end{figure}

Let us examine if 
the two-point function $G_2^{(w)}(L)$
allows the double scaling limit.
In Fig. \ref{opn2ptnr}, we plot $G_2^{(w)}(L)$
against $l=aL$ for $\beta/N=1.0/32$ with $N=32,48,64,96$ 
and $128$.
No scaling behavior is observed.
However, as is the case with the double scaling limit in 
the old-fashioned matrix models \cite{DSL},
we might have to 
admit the wave function renormalization.
Since the Fig. \ref{opn2ptnr} is a log-log plot,
the wave function renormalization amounts to shifting
the curves relatively along the vertical axis.
Fig. \ref{opn2ptr} shows the result for the 
renormalized two-point function
\begin{equation}
  \label{wfr2pt}
  \widetilde{G}_2^{(w)}(L) = \beta^{-0.65} 
G_2^{(w)} (L).
\end{equation}
One can see a clear scaling behavior in the intermediate region
of $l$.
The scaling function is a power of $l$ for large $l$,
and the power is 0.80(2).
The finite $N$ effect can be seen as
a saturation of the power-law behavior
in the large $l$ region,
which shows up at $L \simeq N$.
This is to be compared with the 
case with $N^{-1} G_1^{(n)}(L)$ in Fig. \ref{dblwlp},
where the deviation from the scaling behavior 
in the large $l$ region shows up at $L \simeq N/8$.
This discrepancy is not strange 
since the meaning of $L$ depends on which of the two
operators $O_{\alpha\beta}^{(n)}(L)$ and $O_{\alpha}^{(w)}(L)$
one considers.

\begin{figure}[p]
\begin{center}
\leavevmode\psfig{figure=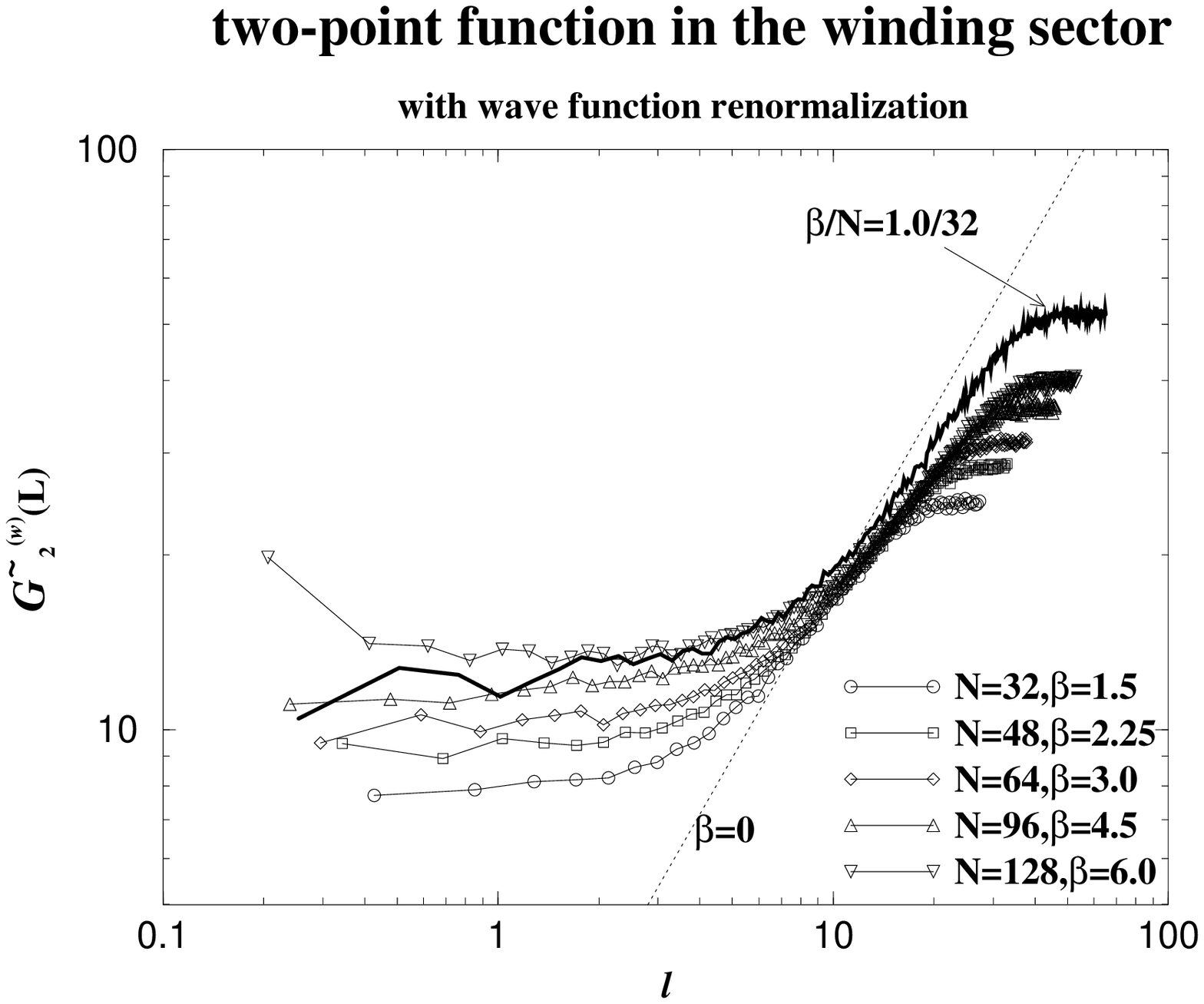,width=16cm,height=12.8cm}
\end{center}
\caption[open loop 2-point function with the wave function renormalization]{
The two-point function in the winding sector 
with the wave function renormalization
$\widetilde{G}_2^{(w)}(L)$ is plotted against $l=aL$
for sets of $\beta$ and $N$ with $\beta/N=1.5/32$.
Each symbol represents the data for $N=32$ (circles), $N=48$ (squares),
$N=64$ (diamonds), $N=96$ (triangles) and $N=128$ (inverse triangles).
The result for $\beta /N=1.0/32$ with $N=128$ are replotted by the heavy line.
The dotted line is a line with slope 1, which represents the 
result for $\beta=0$ and $N=\infty$.}
%strong coupling limit in the large $N$ limit.
\label{opn2ptr2}
\end{figure}

%The discrepancy from the scaling behavior in the large $l$ region
%starts to occur when the number of link variables in the trace is
%around $N$.

Unlike in Fig. \ref{dblwlp},
there is a deviation from the scaling behavior also in the 
small $l$ region.
The tendency of convergence to a scaling function in this region
is not clearly seen up to $N=128$.
The fact that such a deviation does not appear for 
$N^{-1} G_1^{(n)}(L)$ 
in Fig. \ref{dblwlp} might be rather considered
as an accidental property of
the Wilson loop in the non-winding sector.

We can see the scaling behavior for different $\beta/N$ as well.
In Fig. \ref{opn2ptr2},
we plot the renormalized two-point function
$\widetilde{G}_2^{(w)}(L)$
against $l=aL$
for $\beta/N=1.5/32$ with $N=32,48,64,96$ and $128$.
The behavior is qualitatively the same as that of
Fig. \ref{opn2ptr}.
Note that this result shows that
the wave function renormalization does not
depend on $g_{str}=\beta/N$, which is to be expected.
The power of the scaling function in the large $l$ region
is 0.72(3).

In order to understand the power-law behavior,
we calculate
$G_2^{(w)}(L)$ for $\beta = 0$ in the large $N$ limit analytically
through Schwinger-Dyson equation
\footnote{We thank H. Kawai for suggesting the use of 
Schwinger-Dyson equation for this purpose.}.
We consider the quantity
$\langle \mbox{tr}(t^a U_\mu^L) \mbox{tr}(U_\mu^{\dag L}) \rangle$,
where $t^a$ ($a=1,\cdots,N^2$) denote the generators of the U($N$) group.
By changing the variable of integration as
$U_\mu \rightarrow (1- i \epsilon t^a) U_\mu$,
we obtain the following identity.
\beq
  \label{SD-opn2pt}
\sum_{M=1}^{L} 
    \langle 
      \mbox{tr}(t^a U_\mu^{L-M} t^a U_\mu^{M}) 
      \mbox{tr}(U_\mu^{\dag L}) \rangle 
    - L \langle 
      \mbox{tr}(t^a U_\mu^L)
      \mbox{tr}(t^a U_\mu^{\dag L}) \rangle = 0
\eeq
Taking a summation over $a=1,\cdots,N^2$, 
we obtain
\beq
      N \langle \mbox{tr}(U_\mu^L) \mbox{tr}(U_\mu^{\dag L}) \rangle
      + \sum_{M=1}^{L-1} \langle 
        \mbox{tr}(U_\mu^{L-M}) \mbox{tr}(U_\mu^M) \mbox{tr}(U_\mu^{\dag L})
         \rangle
      - N L = 0,
\eeq
where we have used
the identity $\sum_a (t^a)_{ij} (t^a)_{kl} = \delta_{il} \delta_{jk} $.
Since the three-point function 
$\langle \mbox{tr}(U_\mu^{L-M}) \mbox{tr}(U_\mu^M)
\mbox{tr}(U_\mu^{\dag L}) \rangle$ is at most ${\cal O}(1/N)$ 
as we explain later, we obtain
\begin{equation}
  \label{2ptb0}
G_2^{(w)}(L)
=  \langle \mbox{tr}(U_\mu^L) \mbox{tr}(U_\mu^{\dag L}) \rangle = L.
\end{equation}
We have checked this result explicitly by Monte Carlo simulation.
It is natural to expect 
that in the planar limit the two-point function in the large $l$ region 
can be described by the strong coupling limit obtained above.
The fact that the power of $l$ in the scaling function
approaches one for decreasing $g_{str}=\beta/N$ can be naturally 
understood in this way.

Note that the wave function renormalization was actually necessary for 
$G_1^{(n)}(L)$ as well, 
since we had to multiply it by $N^{-1}$ when we see the scaling behavior.
In this sense, 
we should say that the renormalized one-point function is given
by $\widetilde{G}_1^{(n)}(L) = \beta^{-1} G_1^{(n)}(L)$.
Thus, although the result for the two-point function of the Wilson loops
in the winding sector supports to some extent
our conjecture concerning the existence of the double scaling limit,
we encounter a new problem:
why 
the wave function renormalization
for the one-point function $G_1^{(n)}(L)$
is not equal to that for the two-point function
$G_2^{(w)}(L)$.

\begin{figure}[p]
\begin{center}
\leavevmode\psfig{figure=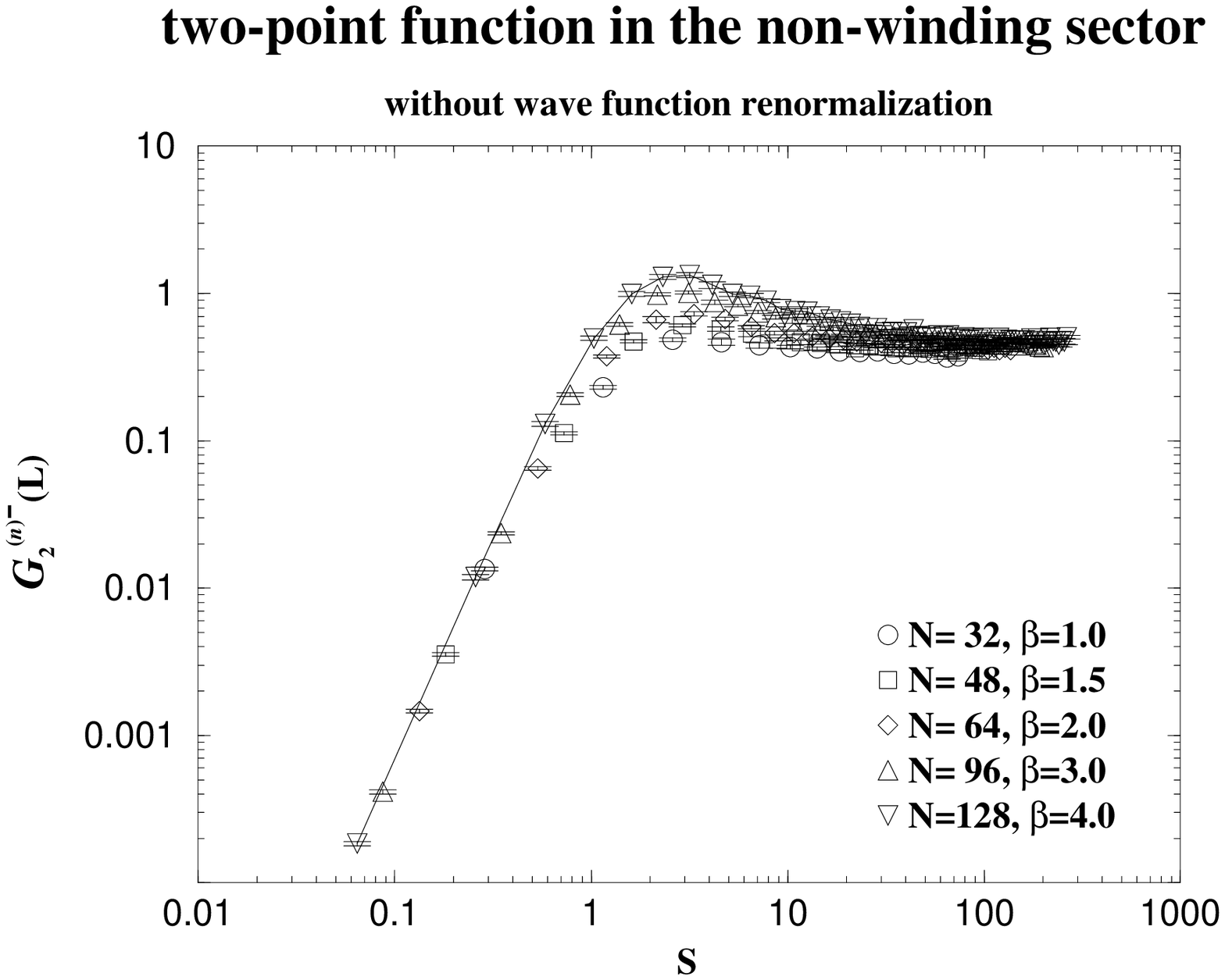,width=16cm,height=12.8cm}
\end{center}
\caption[imaginary part of the Wilson loop 2-point function]{
The two-point function in the non-winding sector 
without the wave function renormalization
$G_2^{(n)-}(L)$ is plotted against $S=(aL)^2$
for sets of $\beta$ and $N$ with $\beta/N=1.0/32$.
Each symbol represents the data for $N=32$ (circles), $N=48$ (squares),
$N=64$ (diamonds), $N=96$ (triangles) and $N=128$ (inverse triangles).
The solid line, which connects the data for $N=128$,
is drawn to guide the eye.
}
\label{wlp2ptnren}
\end{figure}

\begin{figure}[p]
\begin{center}
\leavevmode\psfig{figure=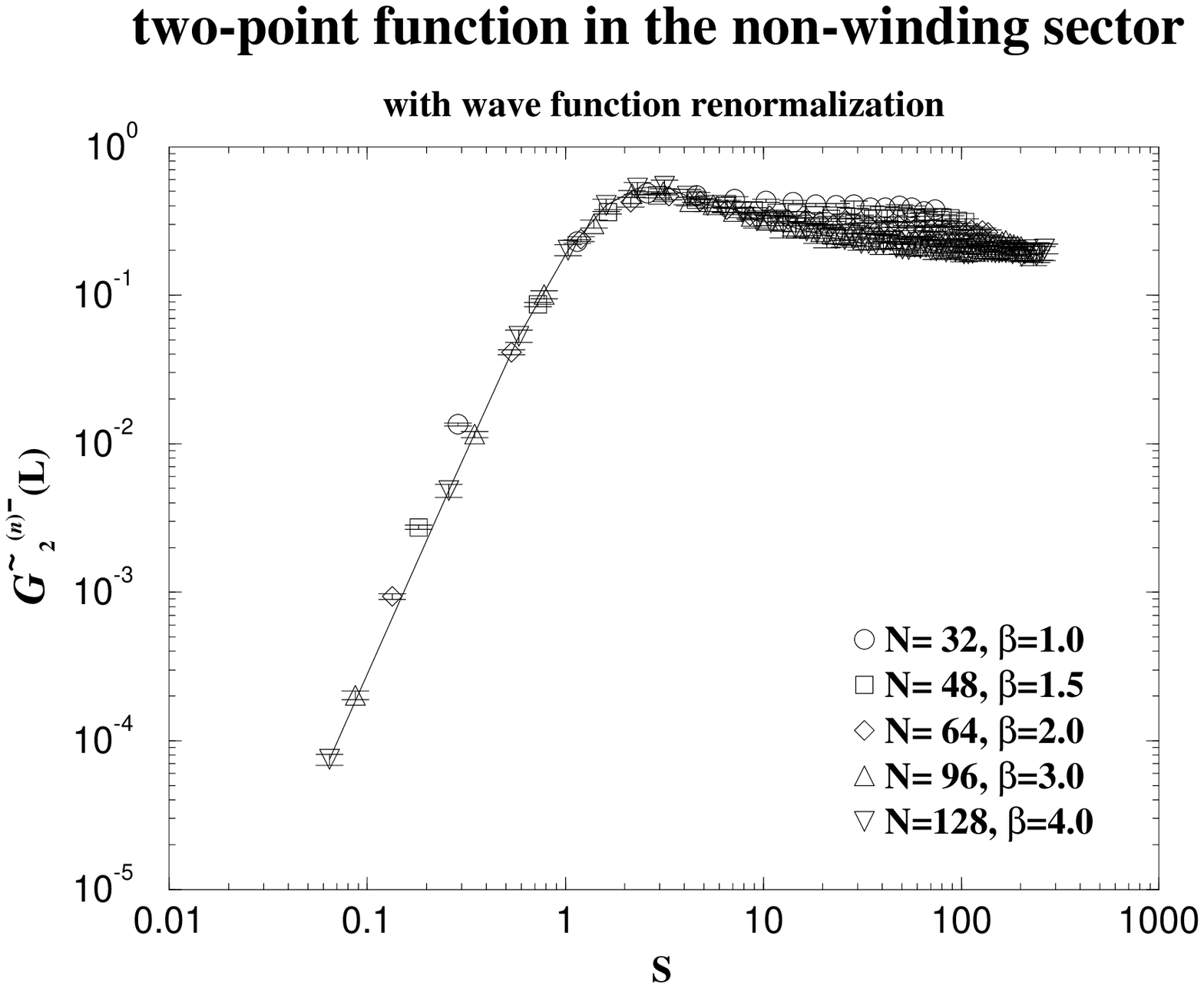,width=16cm,height=12.8cm}
\end{center}
\caption[imaginary part of the Wilson loop 2-point function]{
The two-point function in the non-winding sector 
with the wave function renormalization
$\widetilde{G}_2^{(n)-}(L)$ is plotted against $S=(aL)^2$
for sets of $\beta$ and $N$ with $\beta/N=1.0/32$.
Each symbol represents the data for $N=32$ (circles), $N=48$ (squares),
$N=64$ (diamonds), $N=96$ (triangles) and $N=128$ (inverse triangles).
The solid line, which connects the data for $N=128$,
is drawn to guide the eye.
}
\label{wlp2ptimag}
\end{figure}

In order to clarify this problem, 
we consider the two-point function 
of Wilson loops in the non-winding sector defined by
$\langle O^{(n)}_{12}(L) O^{(n)}_{12}(L) \rangle$ and
$\langle O^{(n)}_{12}(L) O^{(n)}_{21}(L) \rangle$.
These observables are ${\cal O}(N^2)$ quantity,
but the leading term is given by the disconnected part
$\langle O^{(n)}_{12}(L)\rangle \langle O^{(n)}_{12}(L) \rangle$
and
$\langle O^{(n)}_{12}(L) \rangle \langle O^{(n)}_{21}(L) \rangle$,
respectively, due to the factorization.
The connected two-point function can be defined by
\beqa
  \label{subtract}
\langle O^{(n)}_{12}(L) O^{(n)}_{12}(L) \rangle_c
&=&
\langle O^{(n)}_{12}(L) O^{(n)}_{12}(L) \rangle 
- \langle O^{(n)}_{12}(L)\rangle \langle O^{(n)}_{12}(L) \rangle \\
\langle O^{(n)}_{12}(L) O^{(n)}_{21}(L) \rangle_c
&=&
\langle O^{(n)}_{12}(L) O^{(n)}_{21}(L) \rangle
- \langle O^{(n)}_{12}(L) \rangle \langle O^{(n)}_{21}(L) \rangle,
\eeqa
which are ${\cal O}(1)$ quantities.

Since $\langle O^{(n)}_{12}(L) \rangle =
\langle O^{(n)}_{21}(L) \rangle \in {\cal R}$,
as is mentioned in the previous section,
we can avoid the subtraction
by considering the difference of the two-point functions.
\beqa
  \label{2pt-WLP}
G_2^{(n)-}(L) &\equiv&
\frac{1}{2}
\left\{ 
\langle O^{(n)}_{12}(L) O^{(n)}_{21}(L) \rangle_c 
- 
\langle O^{(n)}_{12}(L) O^{(n)}_{12}(L) \rangle_c
\right\} \\
&=& \frac{1}{2}
\left\{ 
\langle O^{(n)}_{12}(L) O^{(n)}_{21}(L) \rangle 
- 
\langle O^{(n)}_{12}(L) O^{(n)}_{12}(L) \rangle 
\right\} \\
&=&  \langle (\mbox{Im}O^{(n)}_{12}(L) )^2 \rangle .
\eeqa
We have used the fact that
$\langle O^{(n)}_{12}(L) O^{(n)}_{12}(L) \rangle =
\langle O^{(n)}_{12}(L) O^{(n)}_{21}(L) \rangle \in {\cal R}$
in the third equality.
In Fig. \ref{wlp2ptnren} we plot the $G_2^{(n)-}(L)$ against $S=(aL)^2$
for $\beta/N=1.0/32$ with
$N=32,48,64,96$ and $128$.
There seems to be no scaling behavior.
Let us assume the same wave function renormalization as 
the one for $G_2^{(w)}(L)$ in (\ref{wfr2pt}).
In Fig. \ref{wlp2ptimag}, 
we plot $\widetilde{G}_2^{(n)-}(L) = \beta^{-0.65} G_2^{(n)-}(L)$
for $\beta/N=1.0/32$ with $N=32,48,64,96$ and $128$.
We find a clear scaling behavior.
The discrepancy in the large $S$ region starts from
$L\simeq N/8$, which is the same as with $\widetilde{G}_1^{(n)}(L)$.
This suggests that the operator $O_{\alpha\beta}^{(n)}(L)$ 
gives a finite $N$ effect to the correlation function
which includes it for $L \simeq N/8$ in general.
It is natural to expect that this is the case also with
the operator $O_{\alpha}^{(w)}(L)$ with $L \simeq N$,
which we exploit in discussing the scaling behavior of the
three-point function of Wilson loops in the winding sector later.
The scaling seems to extend to the small $l$ region
except for the data that corresponds to 
$L=1$.
This is consistent with our previous observation that the 
Wilson loops in the non-winding sector do not
suffer from finite $N$ effects
in the small $l$ region, unlike those in the winding sector.

Similarly, we can define
\beqa
G_2^{(n)+}(L) &\equiv&
\frac{1}{2}
\left\{\langle O^{(n)}_{12}(L) O^{(n)}_{12}(L) \rangle_c
+ \langle O^{(n)}_{12}(L) O^{(n)}_{21}(L) \rangle_c \right\} \\
&=&  \langle (\mbox{Re}O^{(n)}_{12}(L) )^2 \rangle 
  - \langle O^{(n)}_{12}(L) \rangle ^2 .
\eeqa
The result for 
$\widetilde{G}_2^{(n)+}(L)= \beta^{-0.65} G_2^{(n)+}(L)$ also 
shows the scaling behavior, which is qualitatively the same as
in Fig. \ref{wlp2ptimag}, albeit larger error bars 
due to the subtraction.

As in the case of $G_2^{(w)}(L)$,
we can calculate $G_2^{(n)\pm}(L)$ for $\beta = 0$
in the large $N$ limit
by using the Schwinger-Dyson equation
and obtain $G_2^{(n)\pm}(L)=1/2$.
This suggests that in the planar limit, the 
asymptotic behavior of the scaling function of
$G_2^{(n)\pm}(L)$ for large $l$ is constant,
which is consistent with our results.

From these observations, we conclude that 
only the one-point function is exceptional
concerning the wave function renormalization.
Note that the wave function renormalization 
of $G_1^{(n)}(L)$ is, in some sense, kinematically
constrained by the fact that 
$G_1^{(n)}(0) = N$.
The non-vanishing one-point function can be 
considered as the background of
the corresponding string field 
and the exceptional behavior is not very unnatural.

\begin{figure}[p]
\begin{center}
\leavevmode\psfig{figure=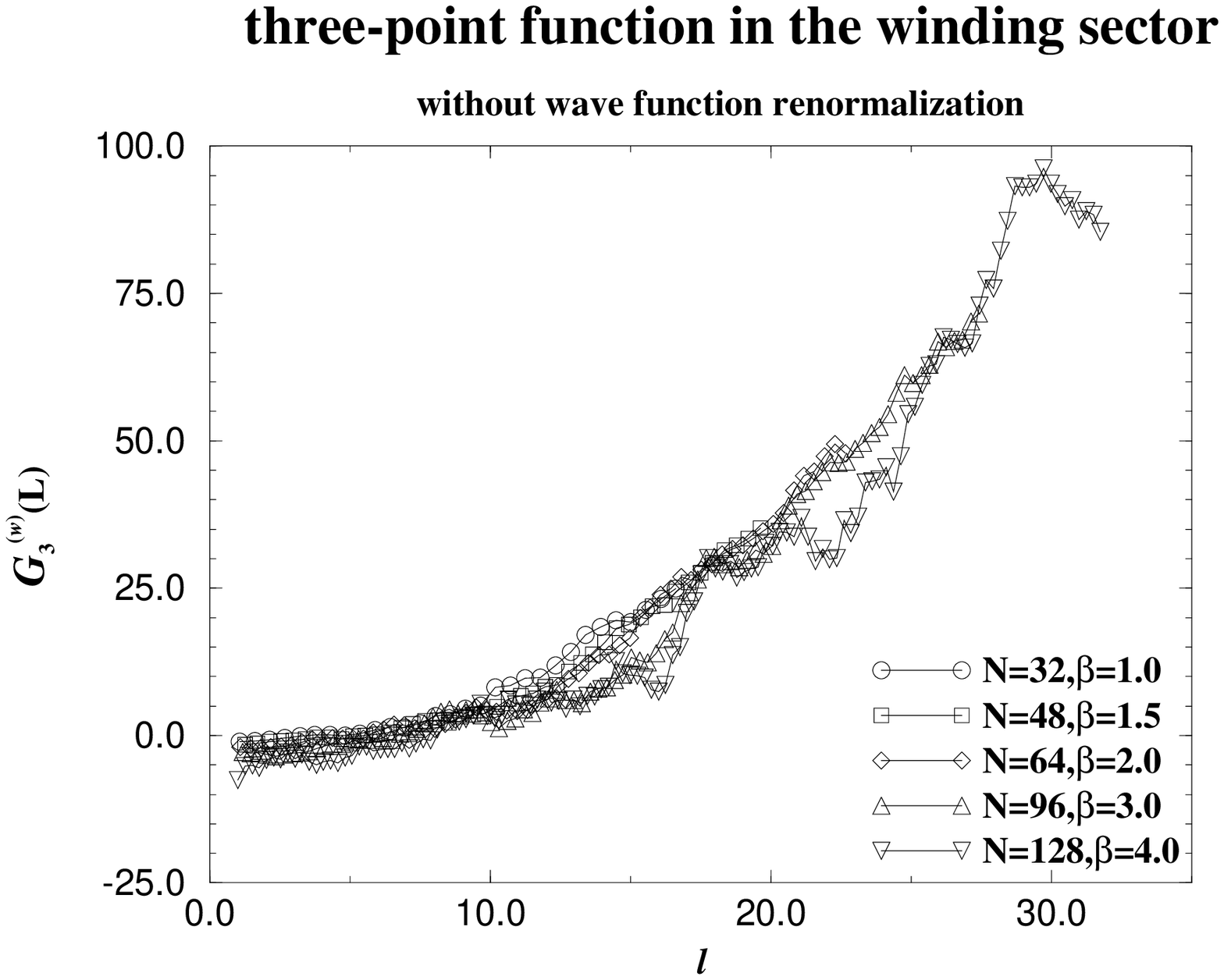,width=16cm,height=12.8cm}
\end{center}
\caption[open loop 3pt-function without wave function renormalization]{
The three-point function in the winding sector 
without the wave function renormalization
$G_3^{(w)}(L)$ is plotted against $l=aL$
for sets of $\beta$ and $N$ with $\beta/N=1.0/32$.
Each symbol represents the data for $N=32$ (circles), $N=48$ (squares),
$N=64$ (diamonds), $N=96$ (triangles) and $N=128$ (inverse triangles).}
\label{3ptnorenorm}
\end{figure}

\begin{figure}[p]
\begin{center}
\leavevmode\psfig{figure=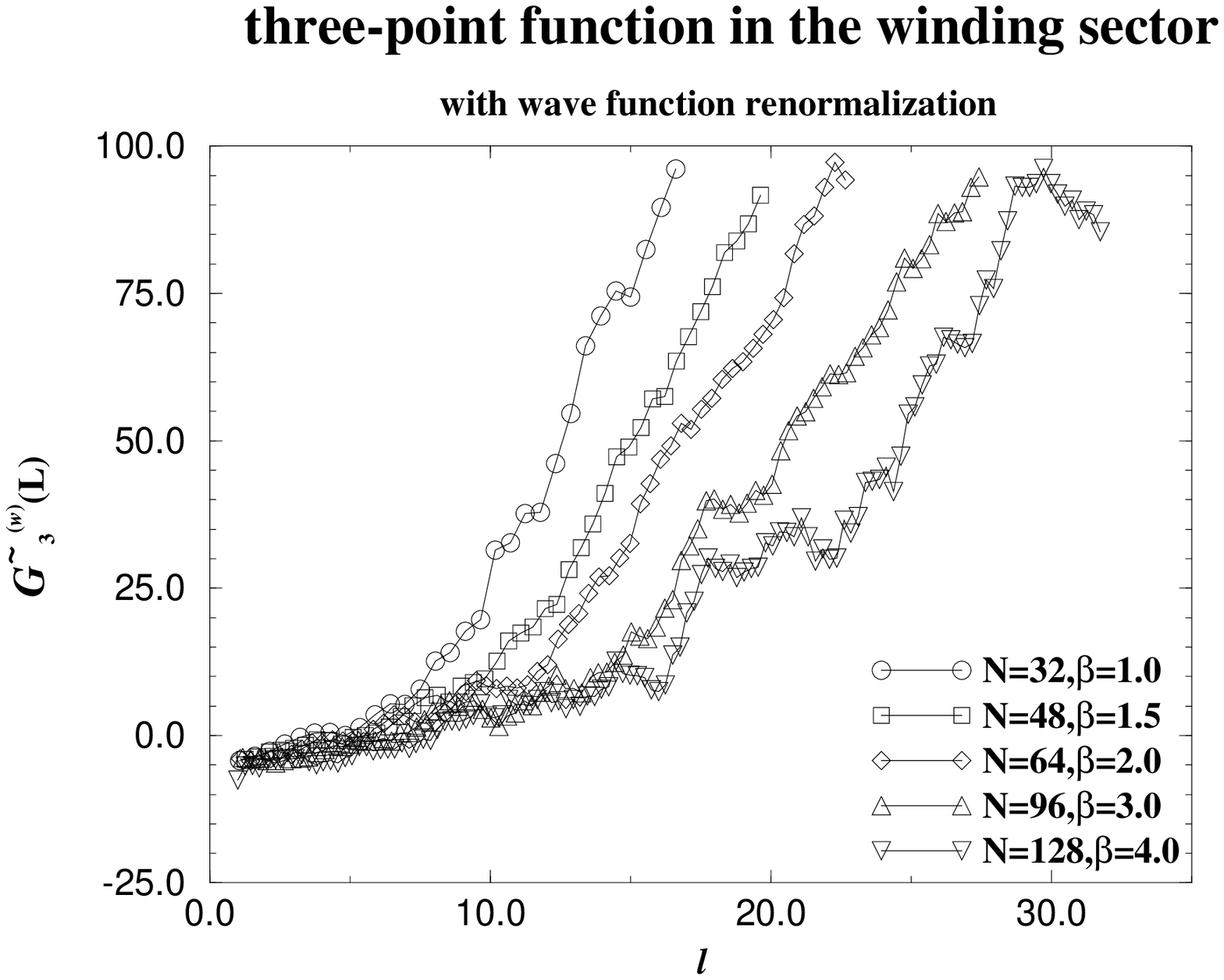,width=16cm,height=12.8cm}
\end{center}
\caption[open loop 3pt-function with wave function renormalization]{
The three-point function in the winding sector 
with the wave function renormalization
$\widetilde{G}_3^{(w)}(L)$ is plotted against $l=aL$
for sets of $\beta$ and $N$ with $\beta/N=1.0/32$.
Each symbol represents the data for $N=32$ (circles), $N=48$ (squares),
$N=64$ (diamonds), $N=96$ (triangles) and $N=128$ (inverse triangles).}
\label{3ptrenorm}
\end{figure}

As a further check of the existence of the double scaling limit,
we consider the three-point function of Wilson loops
in the winding sector defined by
\begin{equation}
  \label{3pt-OPN}
  G_3^{(w)}(L) = 
   \langle O_\mu ^{(w)}(L) O_\mu ^{(w)}(L)  O_{-\mu} ^{(w)}(2L)
\rangle.
\end{equation}
Note that the right hand side does not depend on $\mu$,
and we average over $\mu=\pm 1, \pm 2$ in the measurement 
to order to increase the statistics.
The three-point function is naively 
${\cal O}(N^3)$ quantity,
but the leading term
$\langle O_\mu ^{(w)}(L)\rangle\langle O_\mu ^{(w)}(L) 
\rangle \langle O_{-\mu} ^{(w)}(2L) \rangle$
and the subleading term 
$\langle O_\mu ^{(w)}(L) O_\mu ^{(w)}(L) 
\rangle \langle O_{-\mu} ^{(w)}(2L) \rangle$ {\it etc.},
which are of order ${\cal O}(N)$, vanish
due to the U(1)$^2$ symmetry, and it is actually
${\cal O}(1/N)$ quantity, which can be regarded as the 
connected three-point function.

%We measure the three-point function $G_3^{(w)}(L)$.
In Fig. \ref{3ptnorenorm} and Fig. \ref{3ptrenorm},
we plot
$ G_3^{(w)}(L)$ and
$ \widetilde{G}_3^{(w)}(L) = 
(\beta/4.0)^{-0.65 \times 3/2} G_3^{(w)}(L)$,
respectively, against $l=aL$ for 
$\beta/N=1.0/32$ with $N=32,48,64,96$ and $128$.
Measurements have been done every 10 sweeps.
Each point is an average over 4000 configurations
for $N=32\sim96$, and 
over 2000 configurations for $N=128$.
Note that since the data can be both positive and negative,
we cannot make a log plot, though we should do so 
when we discuss scaling behaviors.
In order to avoid a possible unfair comparison of the figures,
we have defined the $\widetilde{G}_3^{(w)}(L)$ 
so that the data for $N=128$ look the same as those for
$ G_3^{(w)}(L)$.
Although the data are too noisy to confirm the scaling
behavior exclusively 
with the particular wave function renormalization
assumed here, the renormalized data are at least 
consistent with the scaling behavior in some region of $l$.
Furthermore, the deviation from the scaling
in the large $l$ region shows up at $2 L \simeq N$,
which is in agreement with the finite $N$ effects seen for 
the two-point function of the same operator, namely 
$\widetilde{G}_2^{(w)}(L)$.
This is not the case with the data without the wave function renormalization.
We therefore conclude that the data for the three-point function
also support the existence of the double scaling limit with 
the universal wave function renormalization.

\section{Conclusion and discussion}
\setcounter{equation}{0}
\label{sec:conc}

In this paper, we studied the EK model as a toy model of the 
IIB matrix model.
In the planar limit,
the one-point function of the Wilson loops in the non-winding sector
gives the expectation value of 
Wilson loops in the large $N$ lattice gauge theory in two dimensions.
Using the exact results in the 
2D large $N$ lattice gauge theory,
we fixed how to send $a$ to 0 as we send $\beta$ to infinity as
\beq
a = \sqrt{-\log \left( 1-\frac{1}{4\beta} \right) }.
\eeq

In the planar limit, $N$ is sent to infinity before taking the
limit $a\rightarrow 0$ and $\beta \rightarrow \infty$.
We found that there is another way of taking the large $N$ limit,
namely with $\beta/N$ fixed.
We found that the correlation functions of Wilson loops in
the non-winding sector as well as in the winding sector
have nontrivial limits with the wave function renormalization
$\beta^{-0.65/2}$ for each Wilson loop.
The only exception we saw is the one-point function, for which
the wave function renormalization is $\beta^{-1}$.
We consider that this is not so unnatural,
since the one-point function is special in the sense it can be 
considered as the non-vanishing background of the corresponding
string field.

The deviation from the scaling behavior due to the finite $N$ 
is seen in the large $l$ region.
The number of the links for which the Wilson loop operator
gives finite $N$ effects depends on the operator considered.
It is around $N$ for the Wilson loops in the winding sector and
around $N/2$ for the Wilson loops in the non-winding sector.
In either case, the physical scaling region enlarges as 
$\sqrt{N}$ in the large $l$ region.

On the other hand, the situation in the small $l$ region is not so clear.
The deviation from the scaling behavior is seen for 
the two-point functions of Wilson loops in the winding sector.
Convergence to a scaling function seems to be slow.
%Scaling behavior is not clearly seen up to $N=128$.
For the correlation functions of 
the Wilson loops in the non-winding sector, on the other hand,
no significant deviation
from the scaling behavior is seen in the small $l$ region.
We consider this as an accidental property of Wilson loops in 
the non-winding sector.

The data for the $n$-point functions with $n\ge 3$ are unfortunately
too noisy to confirm the scaling behavior for fixed $\beta/N$.
We found, however, that the data for
the three-point function of Wilson loops
in the winding sector with the same wave function renormalization
are consistent with the scaling.
Furthermore, the deviation from the scaling
in the large $l$ region due to the finite $N$ effects appears in a manner
which is consistent with the finite $N$ effects seen for 
the two-point function of the same operator.

We calculated the two-point functions 
for $\beta = 0$ in the large $N$ limit
by using the Schwinger-Dyson equation.
The result seems to describe the large $l$ behavior 
of the corresponding quantity in the planar limit.
We can calculate the connected $n$-point functions in general,
which are order ${\cal O}(1/N^{(n-2)})$ quantities.
We find that the coefficient of $1/N^{(n-2)}$ is zero 
for $\beta = 0$ in the large $N$ limit.
This suggests that the $n$-point functions with $n\ge 3$
in the planar limit goes to zero in the large $l$ region.
One might fear that the string theory constructed through the double 
scaling limit of the EK model is a free theory.
Our data suggest, however, that
the three-point function 
is not constantly zero at least when $g_{str} \ne 0$.

A technically important comment is that while the old-fashioned
matrix model, 
whose action is unbounded from below,
becomes well-defined only in the large $N$ limit,
the EK model as well as the IIB matrix model
is completely well-defined even for a finite $N$.
This makes these models much easier to
study numerically than the old-fashioned matrix model.

To summarize, 
our numerical results suggest strongly that
the double scaling limit of the EK model
can be taken by sending $N$ to the infinity 
with the following combinations of parameters fixed.
\beqa
  \label{dlscombiA}
  \beta/N & \sim &  g_{str}  \\
  N a^2 & \sim & ( \alpha' g_{str})^{-1},
  \label{dlscombiB}
\eeqa
where $g_{str}$ is the string coupling constant and $\alpha'$ is the
string tension.

It is intriguing to ask what is the string theory constructed through
the double scaling limit of the EK model.
We interpreted the model as a string theory through
the T-duality transformation as is done in the IIB matrix model.
The target space is therefore 
two-dimensional space time compactified to a torus
of size $\sim 1/a$, which goes to infinity as $a\rightarrow 0$.
The theory must be different from the $c=1$ matrix model interpreted
as 2D string theory with linear dilaton background, 
since we have at least the discrete rotational invariance
in the EK model.
Note also that although the target space is two-dimensional,
the dynamical degrees of freedom of the link variables
are actually absorbed by the gauge degrees of freedom.
It might be that this theory cannot be described
as 2D gravity coupled to some conformal matter.

An important difference between 
the EK model and the IIB matrix model is
that the dynamical variables in the former are the unitary matrices,
while those in the latter are the hermitian matrices.
Considering the interpretation of the models as string theories
using the T-duality transformation, 
it seems to be more natural
to formulate the matrix models in terms of unitary matrices.
Constructing such a formulation for type IIB superstring 
has been considered in Ref. \cite{NaoNisi}.
The problem corresponding to the fermion doublers in the lattice chiral
gauge theory has been overcome by the use of the overlap formalism,
but the remaining problem is whether one can recover the 
supersymmetry in the large $N$ limit with the proposed model
or with fine-tuning of the coefficients of some possible counterterms.
On the other hand, one could consider 
a hermitian matrix version of the two-dimensional 
EK model, with the action
$ S = - \frac{1}{4g^2} \tr [A_\mu, A_\nu]^2 $,
but the theory is not well defined \cite{SuyamaTsuchiya}.
We can make the theory well defined, for example,
by adding a term like
$m^2 (\tr A_\mu ^2 - \frac{1}{N}(\tr A_\mu)^2)$,
which preserves the U(1)$^2$ symmetry : $A_\mu \rightarrow
A_\mu + \alpha_\mu$.
However, the U(1)$^2$ symmetry is then spontaneously broken
and the model must be in a universality class
other than the one corresponding to the two-dimensional EK model.
 
Despite the above subtlety,
the relations (\ref{dlscombiA}) and (\ref{dlscombiB}) 
which specify how one should take
the double scaling limit in the two-dimensional EK model might be naively 
compared with the ones conjectured for the IIB matrix model
\cite{FKKT}; namely,
\beqa
  \label{dlscombi2A}
  g^2 N &\sim& \alpha ^{'2} \\
  N \epsilon^2 &\sim&  g_{str}^{-1}.
  \label{dlscombi2B}
\eeqa
$\epsilon$ is a dimensionless constant which corresponds naively to $a$.
We therefore have $a \sim \epsilon \alpha  ^{'-1/2}$.
This means that the relation (\ref{dlscombi2B})
coincides with the corresponding one in the EK model,
namely (\ref{dlscombiB}).
The coupling constant $g$ can be related to $\beta$ naively
in the following way.
We relate the hermitian matrices $A_\mu$ in the IIB matrix model
to the link variables $U_\mu$ in 
the EK model through $U_\mu= \exp (iaA_\mu)$,
and expand the action (\ref{red-action}) in terms of $a$.
We obtain $S = - N\beta a^4 \tr [A_\mu, A_\nu]^2 \cdots$.
In Ref. \cite{FKKT}, the factor in front of the trace
is denoted as $1/4g^2$, which means that $1/g^2 \sim N \beta a^4$.
Therefore, (\ref{dlscombi2A}) corresponds naively to 
fixing $\beta/N^2$ in our notation.

We also note that the density of eigenvalues is constant in 
the double scaling limit.
As we mentioned above, 
the extent of the space time is given $R=\frac{2\pi}{a}$.
Since we have $N$ eigenvalues distributed in the two-dimensional
space time with this extent, the average denstity is given by
\beq
\rho = \frac{N}{R^2}= \frac{N a^2}{(2 \pi)^2},
\eeq
which is constant in the double scaling limit.
This fact is natural from the string theoretical point of view,
since it means that there are only finite dynamical degrees 
of freedom on the average in a finite region of the space time.

The fact that a sensible double scaling limit can be taken for the
large $N$ reduced model of two-dimensional lattice gauge theory is itself
encouraging for the research of nonperturbative formulation of 
superstring theory through the IIB matrix model.
We hope to report on numerical studies of the IIB matrix model
in future publications.

\bigskip
\bigskip

\begin{center} 
\begin{large}
Acknowledgments
\end{large} 
\end{center}
We would like to thank H. Kawai for continuous encouragements
and for stimulating discussions.
The authors are also grateful to M. Fukuma, V.A. Kazakov and A. Tsuchiya 
for valuable comments.
This work is supported by the Supercomputer Project (No.97-18)
of High Energy Accelerator Research Organization (KEK).

\end{document}